\documentclass[twocolumn,showpacs,amsmath,amssymb]{revtex4}
\usepackage[dvips]{graphics}

\begin{document}
\title{Motion of charged particles and quasinormal modes around the magnetically and tidally deformed black hole}
\author{R. A. Konoplya}\email{konoplya_roma@yahoo.com}
\affiliation{DAMTP, Centre for Mathematical Sciences, University of Cambridge, Wilberforce Road, Cambridge CB3 0WA, UK}
\affiliation{Theoretical Astrophysics, Eberhard-Karls University of T\"{u}bingen, T\"{u}bingen 72076, Germany}
\affiliation{Centro de Estudios Cient\'{i}ficos (CECS), Casilla 1469, Valdivia, Chile.}
\author{Yu-Chun Liu}
\affiliation{Theoretical Astrophysics, Eberhard-Karls University of T\"{u}bingen,
Auf der Morgenstelle 10, T\"{u}bingen 72076, Germany and
Theoretical Physics Institute, Friedrich-Schiller University of Jena,
Max-Wien-Platz 1, 07743 Jena, Germany}

\begin{abstract}
Here we consider two phenomena in the vicinity of a black hole deformed by the tidal gravitational force of  surrounding matter and by a strong magnetic field: equatorial motion of charged particles and the decay of a test scalar field. We were able to analyze both phenomena with analytical and simple numerical tools, which was unexpected given the low symmetry of the system. We show that both the tidal gravitational force and the magnetic field strongly enhance the release of the binding energy for the matter spiralling into the black hole.
In the presence of the magnetic field, the left and right handed rotations of charged particles are not equivalent and
for sufficiently large $|q| B$ there are stable anti-Larmor orbits very close to the event horizon, although Larmor orbits are only stable at some distance from the black hole. The larger the tidal force, the closer the innermost stable orbit to the black hole for both types of rotation. It was also  shown that the real oscillation frequencies of the characteristic quasinormal modes are considerably suppressed by the tidal force.
\end{abstract}

\maketitle

\section{Introduction}

Although the interior of black holes is invisible, there are reports on observations of astrophysical black holes owing to their intense interactions with the surrounding matter, which are manifest around galactic black holes interacting with the Active Galactic Nucleus.
Accretion disks create gravitational tidal forces as well as electric and magnetic fields around black holes. Accurate physical descriptions of such processes usually involve complex numerical modeling using supercomputers and present a number of challenges connected to explanations of various observed processes, including for example, the origin of jets around black holes. Here we adopt a more theoretical framework, which, although far from being an
accurate astrophysical description, could be a bridge between the mathematical theory of black holes and the numerical modeling in astrophysics.

An important constituent of a black hole's environment is a magnetic field, which can be as strong as $10^{4}-10^{8} G$ near the black hole \cite{1}. The magnetic fields near a black hole can extract its rotational energy  (Blandford-Znajek effect \cite{2}), affect the quasinormal spectrum \cite{Kokkotas:2010zd} and Hawking radiation \cite{Galtsov-book}, \cite{Kokkotas:2010zd}, and induce an external electric charge on its surface \cite{Galtsov-book}.
At the same time, there is no evidence that the magnetic field could be strong enough to deform the geometry of a black hole significantly. Thus, the most interesting effects associated with the presence of the magnetic field occur on charged particles and fields.
Indeed, the factor which stipulates the Lorentz force during the particle's motion is $q B/\mu$ \cite{Galtsov-book}, where $\mu$ is the mass of the particle and $B$ is the asymptotic value of the magnetic field strength.

Another essential factor representing the black hole environment is a tidal gravitational force produced by all the surrounding matter. Unlike magnetic fields, tidal forces can be strong enough to deform (albeit minimumly) the geometry of a galactic black hole as well as black holes of stellar masses having a star companion.

The perturbative solution to the Einstein-Maxwell equations which allows us to study, the magnetic field and the tidal gravitational force, was found in \cite{Preston-Poisson} by Preston and Poisson with the help of the light-cone gauge formalism \cite{Preston:2006zd}.
They considered a system consisting of a black hole and a mechanical structure (a giant torus or long solenoid) at some distance around it, which was the source of the gravitational tidal force and the strong magnetic field. Thus, the background which we shall study here, the Preston-Poisson space-time, has three parameters: mass $M$, magnetic field $B$ and tidal force $\mathcal{E}$.

We investigated two basic properties of particles and fields in the vicinity of such a magnetically and tidally deformed black hole: the proper oscillation frequencies in the black hole's response to the perturbation, termed \emph{quasinormal modes}, and the motion of the charged particles. The quasinormal modes do not depend on the kind of perturbation but only on the parameters of  space-time and are therefore termed the fingerprints of the space-time.
The quasinormal modes are studied in various disciplines such as black hole physics, gauge/gravity correspondence,
gravitational wave astronomy \cite{QNM-collect}, cosmology \cite{cosmology} and quantum gravity \cite{QNM-collect}.

Particle motion and the quasinormal modes have been studied for black holes immersed in a strong magnetic field \cite{Galtsov-book}, \cite{Kokkotas:2010zd},  \cite{Preti:2009zz}, \cite{Preti:2005bu}, \cite{Aliev:2002nw}, \cite{Aliev:1989wx}, \cite{Esteban}, \cite{Konoplya:2006gg}, \cite{Konoplya:2007yy}, \cite{Konoplya:2008hj}.We previoulsy investigated motion of neutral particles in the vicinity of the Preston-Poisson space-time \cite{Konoplya:2006qr}. Further studies on particle motion, accretion and thermodynamic properties around black holes in a magnetic field was performed in \cite{Kovacs:2011ec}, \cite{Abdujabbarov:2009az}, \cite{Frolov:2010mi}, \cite{Rahimov:2011fv}, \cite{Konoplya:2010ak}. In the presented work we estimate quaisnormal modes of a test scalar field propagating in the background of the Preston-Poisson black hole \cite{Preston-Poisson}, when the magnetic component of the deformation vanishes.
This is a good approximation, since the gravitational tidal force is usually much stronger than the deformation caused by the magnetic field.
In the ekional regime, quasinormal modes usually have a kind of universal behavior for different boson fields, such that the spectrum of the neutral scalar field considered here could reveal some features of the gravitational spectrum. The latter  is much more difficult for analysis due to the inseparability of variables in the perturbation equations.

The motion of charged particles is essentially influenced by the magnetic component. Therefore, in the first part of our work, when considering equatorial motions of charged particles, we have taken both the gravitational tidal force and the space-time deformation due to the magnetic field of the torus into consideration.
The angular variables in the Hamilton-Jacobi equation in the Preston-Poisson space-time can be decoupled only in the equatorial plane, by which our analysis of particle motion is limited. In particular, we show that the tidal gravitational force as well as the magnetic field strongly enhance the release of the binding energy for a particle spiralling into the black hole. The region of stability is also significantly affected by the tidal force and magnetic field and is qualitatively different for left and right handed rotations. We shall discuss these issues in detail in Sec. IV.

The scalar field equation also does not allow for a complete separation of variables in the general case. Nevertheless, we used the fact that the astrophysically motivated low-laying quasinormal modes are "localized" near the peak of the effective potential and therefore "averaged" the tidal force near the peak by its value in the maximum. This allowed us to decouple variables and estimate the dominant quasinormal modes for the first and higher multipoles $\ell \geq 1$.
We have shown that the real oscillation frequencies are considerably suppressed by the tidal force in the vicinity of the magnetic field.

The paper is organized as follows. In Sec. II we summarize some basic relations for the Preston-Poisson space-time. In Sec. III the effective potential for the equatorial motion is deduced on the basis of the Hamilton-Jacobi formalism. Properties of particle's motion are studied in Sec. IV with consideration of various particular cases, such as zero tidal force, zero magnetic field, and the approximation of the neglected "geometric" influence of the magnetic field.  Sec. V is devoted to the decoupling of variables in the test scalar field equation for the Preston-Poisson metric with vanishing magnetic field. In Sec. VI, the angular part of the scalar field equation is studied.  Sec. VII presents calculations of quasinormal modes for the above case. In Sec. VIII  we summarize the obtained results and discuss the questions which remain for further study.

\section{Preston-Poisson metric}

\begin{figure*}
\resizebox{\linewidth}{!}{\includegraphics*{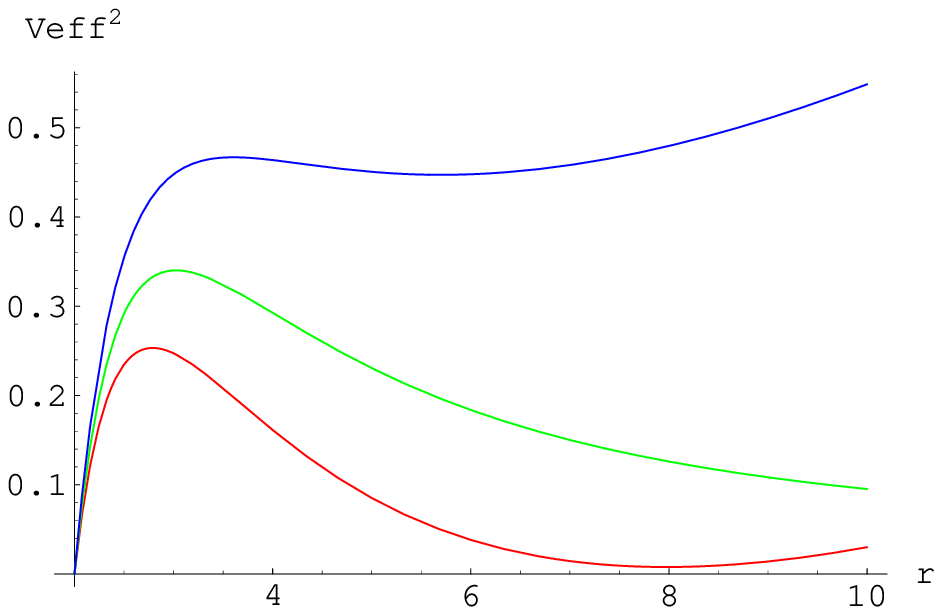}\includegraphics*{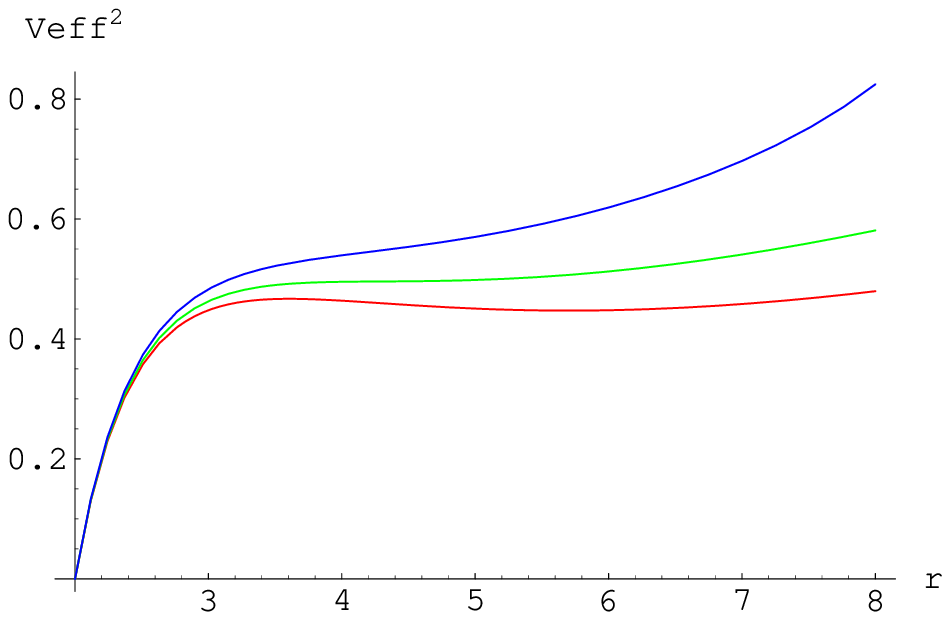}\includegraphics*{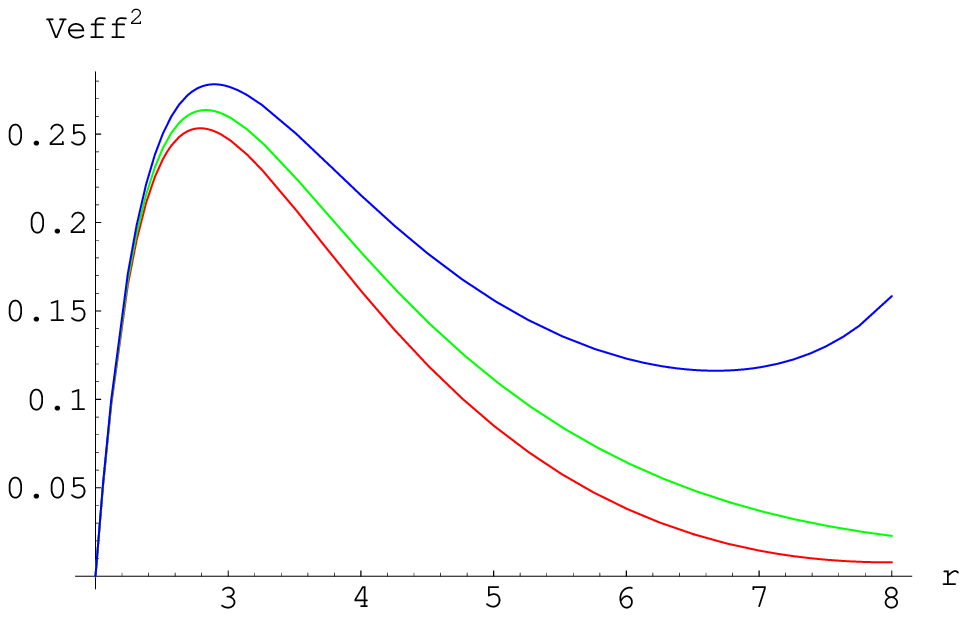}}
\caption{a) Left panel: Effective potentials $U_{eff}^{2}$ for $L = 3$, $\mu=0.1$, $\mathcal{E} = 0.001$, $B=0.01$, $M=1$, $q=-10, 0, +10$ (from  bottom to top), b) middle panel: Effective potentials for $L = 3$, $\mu=0.1$, $\mathcal{E} = 0.001, 0.005, 0.01$ (from bottom to top), $B=0.01$, $M=1$, $q=+10$, c) right panel: Effective potentials for $L = 3$, $\mu=0.1$, $\mathcal{E} = 0.001, 0.005, 0.01$ (from bottom to top), $B=0.01$, $M=1$, $q=-10$}\label{fig0}
\end{figure*}

\begin{figure}
\resizebox{0.8 \linewidth}{!}{\includegraphics*{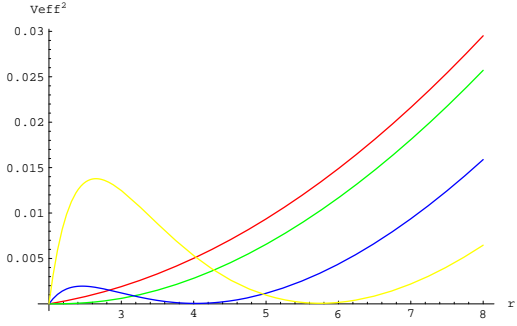}}
\caption{Effective potential $U_{eff}^{2}$ for $L = 0, 0.1, 0.4, 0.8$ (from top to bottom on the right side), $M=1$, $\mu =0.01$, $B=0.01$, $\mathcal{E} = 0.001$, $q=-5$.}\label{fig1}
\end{figure}

The Preston-Poisson space-time describes the system consisting of a large mechanical structure, such as a giant solenoid or a torus which surrounds a black hole and produces an asymptotically uniform magnetic field of strength $B$. The mass of the structure is $M'$ and the radius is of order $\sim a$, while the mass of the black hole is $M$.
The electromagnetic four-vector has the following form,
\begin{equation}
A^{\mu} = \frac{1} {2} B \phi^{\mu},
\end{equation}
where $\phi^{\mu} = (0, 0, 0, 1)$. Further, it is implied that the perturbation created by the magnetic field is small, i.e.
\begin{equation}\label{cond1}
r^2 B^2 \ll 1,
\end{equation}
where $r$ is the distance from the black hole, and only the interior of the mechanical structure is under consideration $r < a$.
For our purposes it is sufficient to study the inner region of the system
which starts at the black hole horizon ($r_h = 2 M$) and ends far from the black hole, still being far from the edges of the torus,
\begin{equation}\label{cond2}
r_h \leq r \ll a.
\end{equation}
The latter condition can always be fulfilled, because the torus is supposed to be situated in the region of the weak gravitational field of the
black hole,
\begin{equation}\label{cond3}
\frac{M}{a} \ll 1.
\end{equation}
As $r^2 B^2 \ll 1$ and $r < a$, it is implied that $a^2 B^2 \ll 1$, though the relative scales of $M/a$ and $a^2 B^2$ can be arbitrary.
First of all, we are interested in the case
\begin{equation}\label{cond4}
M/a \ll a^2 B^2,
\end{equation}
i.e. in the situation when there is an asymptotic region $M \ll r < a$, where the influence of the magnetic field on the space-time geometry is negligible.
The mechanical structure of mass $M'$ produces the gravitational tidal force, parameterized by  $\mathcal{E}$, near the black hole,
\begin{equation}\label{cond5}
\mathcal{E} \sim \frac{M'}{a^3}.
\end{equation}
In the above approach the tidal force can be much larger, of the same order or much smaller than $B^2$.

Preston and Poisson used the light-cone gauge for constructing the perturbed metric, which is adapted  to incoming light cones $v =
constant$ that converge toward $r = 0$. For zero tidal force and magnetic field, $v$ takes its Schwarzschild value
$v = t + r + 2M ln(r/2M - 1)$.
In the ($v$, $r$, $\theta$, $\phi$) coordinates the Preston-Poisson metric has the following form
$$ ds^2 = g_{vv}(r, \theta) dv^2 + 2 dv dr + g_{v \theta}(r,\theta) dv d \theta +  $$
\begin{equation}\label{PPmetric}
g_{\theta \theta}(r,\theta) d \theta^2 +
 g_{\phi \phi}(r,\theta) d\phi^2,
\end{equation}
where
\begin{widetext}
\begin{equation}\label{PPcomponents1}
g_{vv}=-f-\frac{1}{9}B^2r(3r-8M)-\frac{1}{9}B^2(3r^2-14Mr+18M^2)(3\cos^2\theta-1)    +\mathcal{E}(r-2M)^2(3\cos^2\theta-1)+O[B^4,\mathcal{E}^2],
\end{equation}
\begin{equation}\label{PPcomponents2}
g_{vr}=1+O[B^4,\mathcal{E}^2]\text{,}
\end{equation}
\begin{equation}\label{PPcomponents3}
g_{v\theta}=\frac{2}{3}B^2r^2(r-3M)\sin\theta\cos\theta-2\mathcal{E} r^2(r-2M)\sin\theta\cos\theta+O[B^4,\mathcal{E}^2],
\end{equation}
\begin{equation}\label{PPcomponents4}
g_{\theta\theta}=r^2-\frac{2}{9}B^2r^4+\frac{1}{9}B^2r^4(3\cos^2\theta-1)+B^2M^2r^2\sin^2\theta+\mathcal{E} r^2(r^2-2M^2)\sin^2\theta+O[B^4,\mathcal{E}^2],
\end{equation}
\begin{equation}\label{PPcomponents5}
g_{\varphi\varphi}=r^2\sin^2\theta-\frac{2}{9}B^2r^4\sin^2\theta+\frac{1}{9}B^2r^4\sin^2\theta(3\cos^2\theta-1)-B^2M^2r^2\sin^4\theta -\mathcal{E} r^2(r^2-2M^2)\sin^4\theta+O[B^4,\mathcal{E}^2],
\end{equation}
\end{widetext}
and $f(r) = 1- (2 M/r)$.
The above metric was obtained in \cite{Preston-Poisson} by the perturbation of the Enstein-Maxwell equation in orders of $(\mathcal{E}, B^2)$.
The parameter $\mathcal{E}$  is the Weyl curvature, that is, the tidal
gravitational field, of the asymptotic space-time measured
by an observer co-moving with the black hole in the
region $M \ll r\ll 1/B$.
The perturbed event horizon is given by
\begin{equation}
r_{h} = 2 M (1+ \frac{2}{3} M^2 B^2 \sin^2 \theta).
\end{equation}
It is essential
that the event horizon is affected by $B$ and not by $\mathcal{E}$.

Although in the above space-time only the dominant order of the magnetic field is considered, the same relation as for the Ernst black holes takes place
\begin{equation}
B \sim 10^{-21} \frac{M}{M_{\odot}} B_{0},
\end{equation}
where $M$ and $M_{\odot}$ are the mass of a black hole and of the sun respectively and $B_{0}$ is the external magnetic field in units of gauss.
From the above relation one can see that the magnetic field deforming the space-time geometry significantly would be as strong as $B = 10^{12} G$ for galactic black holes with  mass $M \sim 10^{9} M_{\odot}$.

\section{Equations of motion and the effective potential}

The transformations
\begin{equation}
v = t + F(r), \quad F'(r) g_{vv}(r, \theta) + 1 =0
\end{equation}
reduce the metric (\ref{PPmetric}) to the following form
\begin{equation}\label{reduced-metric}
ds^2 = - g_{vv}(r, \theta)^{-1} d r^2 + g_{vv}(r, \theta) dt^2 + g_{\phi\phi}(r, \theta) d\phi^2 + C,
\end{equation}
where $C$ is the part of the metric which vanishes on the equatorial plane $\theta = \pi/2$ as it contains $d \theta$. The metric coefficients $g_{vv}$ and $g_{\phi \phi}$ depend on the radial and angular coordinates $r$ and $\theta$, the black hole mass $M$, magnetic field $B$ and the tidal force parameter $\mathcal{E}$. The above metric is diagonal on the equatorial plane and convenient for further analysis of geodesic motions with the help of the Hamilton-Jacobi formalism.

The Hamilton-Jacobi equations for the time-like and null geodesics in curved space-time are
\begin{equation}\label{H-J}
\frac{1}{2}g^{\mu \nu} \left(\frac{\partial S}{\partial x^{\mu}} - q A_{\mu}\right) \left(\frac{\partial S}{\partial x^{\nu}} - q A_{\nu}\right) = \frac{\partial S}{\partial s}.
\end{equation}
Here $s$ is an invariant affine parameter.
The action is
$$S = -\frac{1}{2} \mu^2 s - E t + L \phi + S_{r}(r) + S_{\theta}(\theta),$$
and $E$ and $L$ are constants of motion, the particle's energy and angular momentum
respectively ($p_{0} = - E$, and $p_{3} = L$) while $p_1 = S_{r}$ and $p_2 = S_{\theta}$ are functions of $r$ and $\theta$ respectively.
In the general case the angular variables in the Hamilton-Jacobi equations cannot be decoupled.
Nevertheless, one can decouple variables for motion in the purely equatorial plane $\theta =\pi/2$.

In the equatorial plane  the Hamilton-Jacobi equation for the metric (\ref{reduced-metric}) takes the form
$$ -g_{vv}^{-1} \left(\frac{\partial S}{\partial t}\right)^{2} + g_{vv} \left(\frac{\partial S}{\partial r}\right)^{2} + g_{\phi \phi}^{-1} \left(\frac{\partial S}{\partial \phi}\right)^{2}  $$
\begin{equation}
+  q^2 g_{\phi \phi}^{-1} A_{\phi}^2 - 2 q A_{\phi} g_{\phi \phi}^{-1} \frac{\partial S}{\partial \phi}  = \mu^2.
\end{equation}

Implying the normalization $\tau = \mu s $, where $\tau$ is the proper time, the first integrals of motion are
$$ - \mu g_{vv} \frac{d t}{d s}= E, \quad \mu \frac{d r}{d s} = \pm \sqrt{(E^2 - U_{eff}^2)}, $$
\begin{equation}
\mu \frac{d \phi}{d s} = \frac{L + \frac{1}{2} B q g_{\phi \phi}}{g_{\phi \phi}}.
\end{equation}

The qualitative analysis of the motion can be made by considering the effective potential
\begin{equation}
U_{eff}^2  =  - \mu^{2} g_{vv} \left(1+ \frac{(L + \frac{1}{2} B q g_{\phi \phi})^2}{\mu^{2} g_{\phi \phi}}\right).
\end{equation}
The effective potential for various parameters is shown in fig. \ref{fig0}. Let us take $M=1$ and consider the situation when, $L \gg B q$ and at the same time $B \ll 1$, $\mathcal{E} \ll 1$, $\mu \ll 1$. Expansion in terms of small $\mu$, $B$ and $|q| B/L$ shows that for sufficiently large values of $L$, the effective potential has a local minimum at

\begin{widetext}
\begin{equation}
r \approx \left(\frac{2 L}{|q| B}\right)^{1/2} \left(1-\left(1 + \frac{q B}{L} \right)\mathcal{E} + \left(\frac{7 q B}{L}^{2}+
 \frac{10 q B}{L} +3 \right) \mathcal{E}^2\right).
\end{equation}
\end{widetext}
For small values of the angular momentum $L$ the effective potential is a monotonically growing function of $r$.
For more general cases, which are not limited by small values of $\mu$, it may be useful to introduce the angular momentum, charge and energy "normalized" by mass:
$$L \rightarrow L/\mu, \quad q  \rightarrow q/\mu, \quad  E  \rightarrow E/\mu.$$
Then, the effective potential can be written as
\begin{equation}\label{general-eff-pot}
V_{eff}^2 = \frac{U_{eff}^2}{\mu^2}  =  - g_{vv} \left(1+ \frac{(L + \frac{1}{2} B q g_{\phi \phi})^2}{ g_{\phi \phi}}\right).
\end{equation}
Further, we shall use this form of the effective potential and the "normalized" values of the energy, angular momentum and charge.

The effective potential, and consequently the motion of particles, is not the same for opposite charges of particles $q$, that is, the left-hand and right-hand rotations are not equivalent due to the presence of the magnetic field: the rotation which corresponds to the Lorentz force directed from the black hole is called the \emph{anti-Larmor} rotation, while the rotation with a Lorentz force directed towards the black hole is the \emph{Larmor} one.  In the next section, we shall show that negative (positive) charge and the positive (negative) angular momentum corresponds to the Lorentz force directed from the black hole (anti-Larmor motion), and, when the signs of the charge and of the momentum coincide, the motion is Larmor.

From fig. \ref{fig0} one can see that the effective potential of positively charged particles is higher than that of the negatively charged ones at all $r$. Thus, the magnetic field
allows negative particles to penetrate the barrier at energies smaller than those for the Schwarzschild black hole.

Now, we are in position to study properties of the equatorial motion of charged particles.

\section{Properties of particle motion}

Here, we shall concentrate on circular orbits of charged, massive particles.
The case of massless and neutral particles was considered, though in a different coordinate system, in \cite{Konoplya:2006qr}.

\subsection{The case where the influence of the magnetic field on the black hole's geometry vanishes}

We shall first neglect the deformation of the black hole geometry due to the magnetic field, that is,
we shall take $B=0$ in formulas
(\ref{PPmetric}) for $g_{vv}$ and $g_{\phi \phi}$ and keep $B \neq 0$ in the generalized derivatives of the Hamilton-Jacobi equations (\ref{H-J}).
This approximation is quite good, taking into account that the magnetic field around the black hole decays quite quickly with $r$, so that
its overall influence on the geometry can indeed can be neglected for small $q B$.

The parameters of the circular orbits can be determined from the requirements:
\begin{equation}
\frac{d r}{d s} =0, \quad \frac{d^2 r}{d s^2} =0.
\end{equation}
These equations mean that on the circular orbits one has
\begin{equation}\label{orbits1}
V_{eff} = E, \quad \frac{d V_{eff}^2}{d r} =0,
\end{equation}
that is, circular orbits take place only on the turning points of the effective potential. Next, let us consider the two particular cases: 1) vanishing tidal force and 2) vanishing magnetic field.

\subsubsection{The case $\mathcal{E} =0$}
In the case of vanishing tidal force, $\mathcal{E} =0$.
Solutions to the system of equations (\ref{orbits1}) allow us to find the values of energy $E$ and momentum $L$ as
a function of the rest of the parameters $B$, $M$, $q$ and coordinate $r$,

\begin{widetext}
\begin{equation}\label{EandL1}
E^2 = \frac{(r-2 M)^2 \left(B^2 q^2 (r-2 M) r^2+2 r-6 M \mp B q \sqrt{r^2 \left(B^2 q^2 r^4+4 M^2 \left(B^2 q^2 r^2-3\right)+M \left(4 r-4 B^2 q^2
   r^3\right)\right)}\right)}{2 r (r-3 M)^2}
\end{equation}
\begin{equation}\label{EandL2}
L=  \frac{\pm \sqrt{r^2 \left(B^2 q^2 r^4+4 M^2 \left(B^2 q^2 r^2-3\right)+M \left(4 r-4 B^2 q^2 r^3\right)\right)} - B M q r^2}{6 M-2 r}.
\end{equation}
\end{widetext}

When $B=0$, the above expressions (\ref{EandL1}, \ref{EandL2}) for the energy and momentum reduce to their Schwarzschild values
\begin{equation}
E^2 = \frac{(r-2 M)^2}{r (r-3 M)},\quad L = \mp \frac{\sqrt{M} r}{\sqrt{r-3 M}}.
\end{equation}

Note that $L$, $E$ and $q$ are "normalized" by mass $\mu$.
Under positive $q$, the upper sign corresponds to the anti-Larmor motion (A), while the lower sign, to the Larmor one (L).
Negative $q$ gives opposite correspondence for Larmor and anti-Larmor rotations.
Indeed, asymptotic behavior at large (far from the black hole) $r$, gives the following asymptotic values for the energy and momentum:
\begin{equation}
E_{A}^2 \rightarrow 1, \quad L_{A} \rightarrow -\frac{1}{2} B q r^2, \qquad r \gg M,
\end{equation}
\begin{equation}
E_{L}^2 \rightarrow  B^2 \left(3 M^2+r^2\right) q^2+1, \qquad r \gg M,
\end{equation}
\begin{equation}
L_{L} \rightarrow \frac{1}{2} B q \left(6 M^2+2 r M+r^2\right), \qquad r \gg M.
\end{equation}

The Larmor motion is a kind of cyclotron rotation in a uniform magnetic field, when
the magnetic field is directed to the center of the orbit.
The Larmor motion does not require a black hole, so that a black hole just perturbs the existing Larmor orbits.
On the contrary to the Larmor rotation, anti-Larmor motion takes place only in the presence of the black hole, when the Lorentz force is directed
outwards from the center.

\subsubsection{The case $B =0$}

The case of vanishing magnetic field is more trivial as the motion of charged particles is qualitatively the same as those of the neutral ones.
The energy and momentum expanded until the first order in $\mathcal{E}$ have the following form:
$$ E^2 = \frac{\mathcal{E} \left(8 M^2-7 r M+2 r^2\right) (r-2 M)^2}{(r-3 M)^2} + $$
\begin{equation}
\frac{(r-2 M)^2}{r (r-3 M)} + O(\mathcal{E}^2),
\end{equation}
$$ L = \pm \frac{\mathcal{E} \left(-6 M^4-2 r M^3+9 r^2 M^2-5 r^3 M+r^4\right) r}{2 \sqrt{M} (r-3 M)^{3/2}} \pm $$
\begin{equation}
\frac{\sqrt{M} r}{\sqrt{r-3 M}} + O(\mathcal{E}^2).
\end{equation}
It is evident that the above expressions reduce to their Schwarzschild values as $\mathcal{E} =0$.

\subsubsection{The case $B \neq 0$ and $\mathcal{E} \neq 0$}

\begin{figure*}
\resizebox{0.7 \linewidth}{!}{\includegraphics*{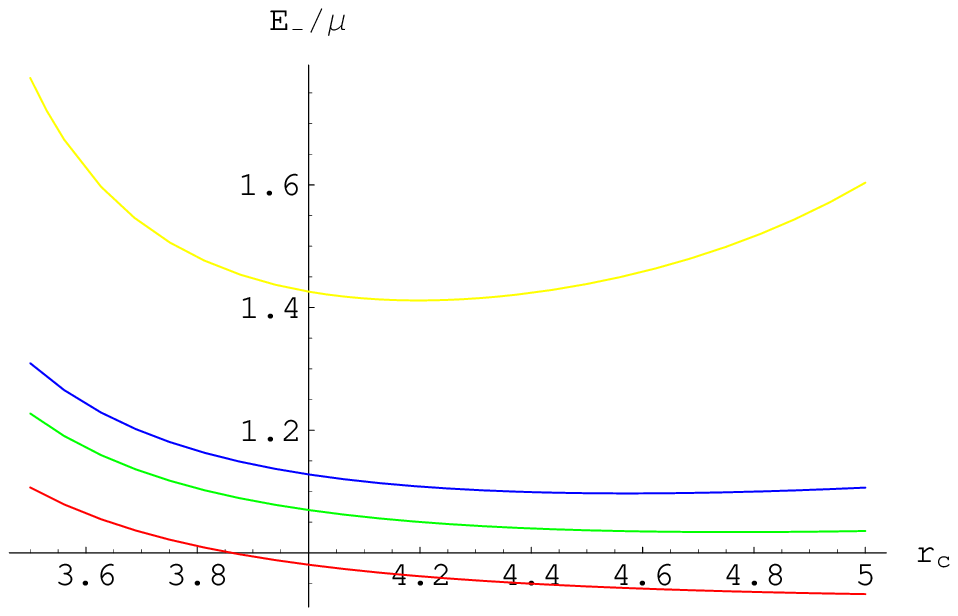}\includegraphics*{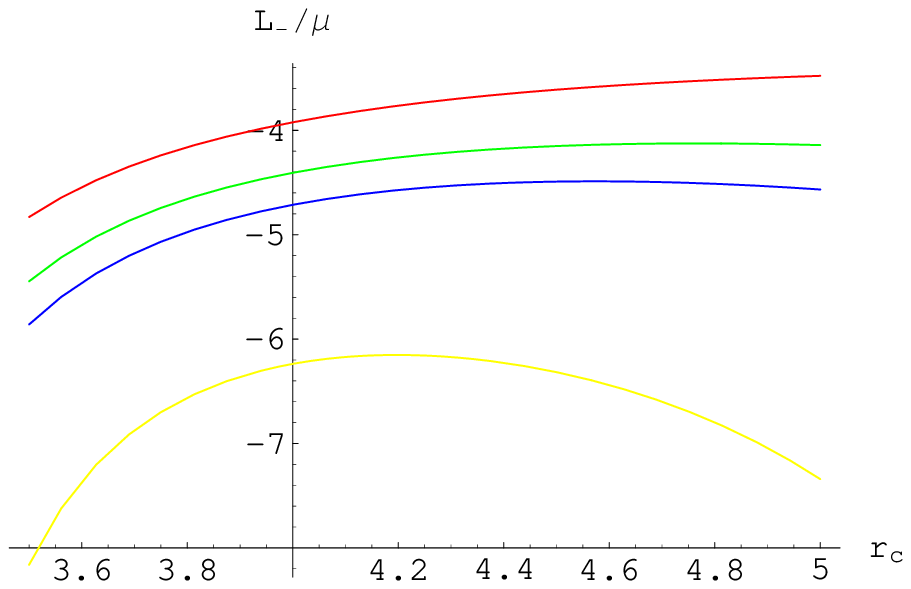}}
\caption{Energy $E_{-}$ (left) and momentum  $L_{-}$ (right) as a function of circular orbit radius for various values of  $\mathcal{E} =0, 1/300, 1/200, 1/100$ (from bottom to top for energy $E_{-}$ and from top to bottom for the negative angular momenta $L_{-}$), $B=1/100$, $M=1$, $q=1$.}\label{figA1}
\end{figure*}

\begin{figure*}
\resizebox{0.7 \linewidth}{!}{\includegraphics*{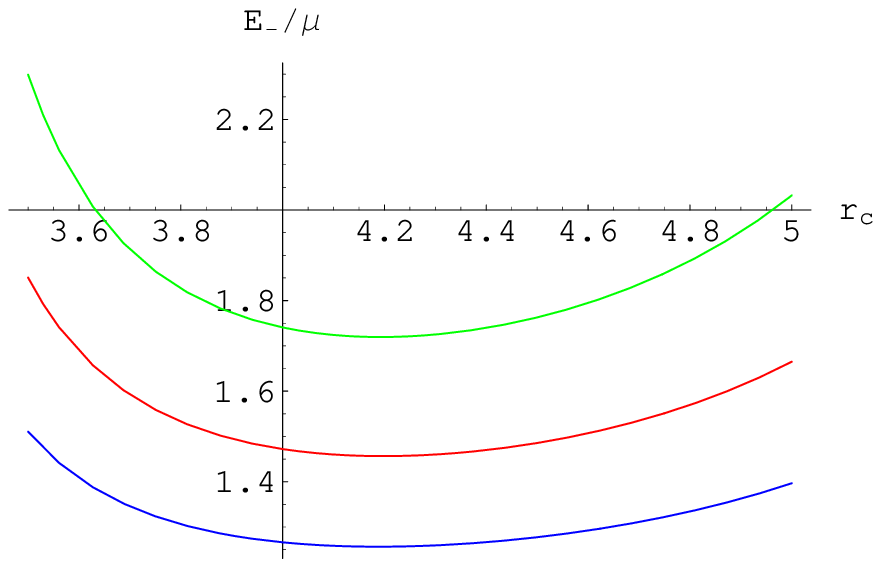}\includegraphics*{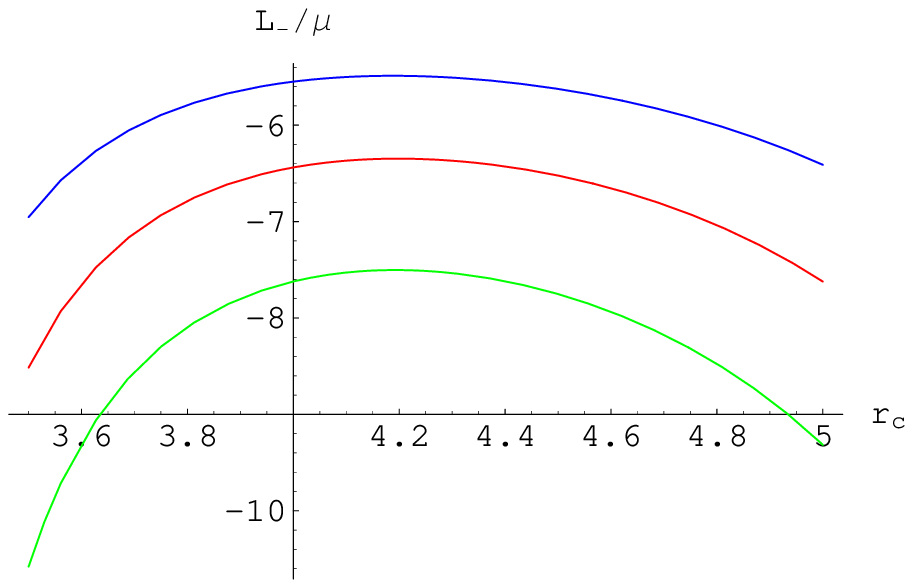}}
\caption{Energy $E_{-}$ (left) and momentum  $L_{-}$ (right) as a function of circular orbit radius for various values of the particle charge $q = -5$(green), $0$(red), $5$ (blue) $\mathcal{E} =1/100$, $B=1/100$, $M=1$.}\label{figA2}
\end{figure*}

\begin{figure*}
\resizebox{0.7 \linewidth}{!}{\includegraphics*{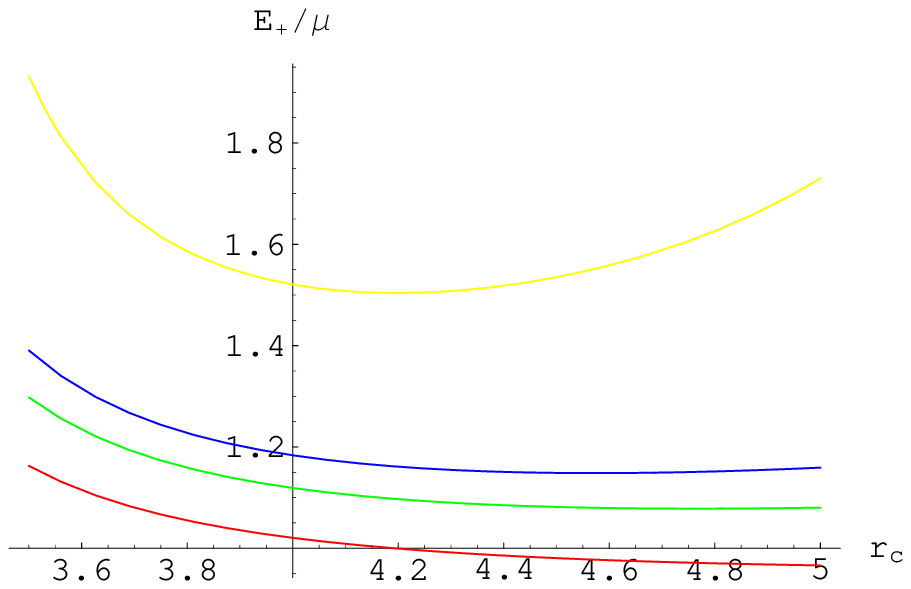}\includegraphics*{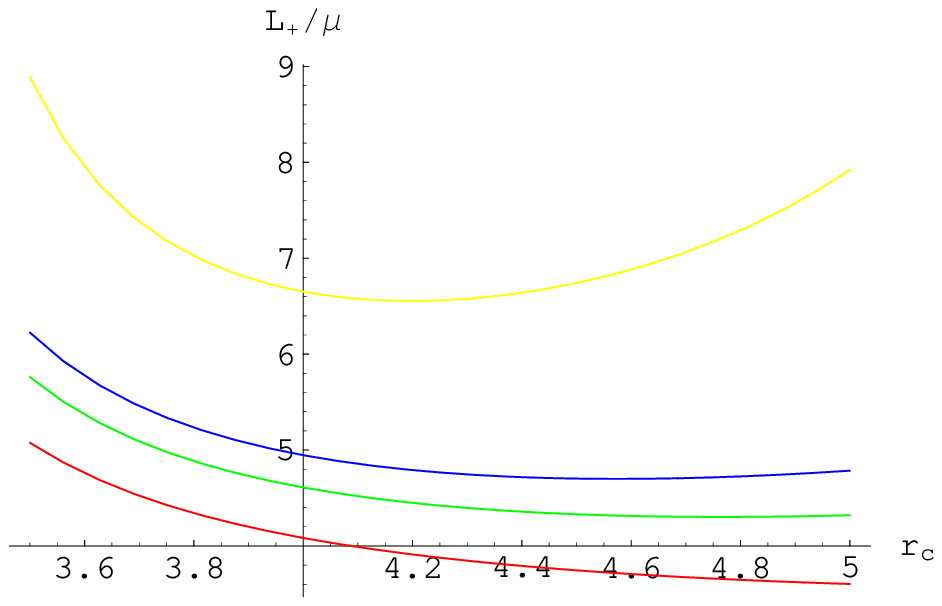}}
\caption{Energy $E_{+}$ (left) and momentum  $L_{+}$ (right) as a function of circular orbit radius for various values of  $\mathcal{E} =0, 1/300, 1/200, 1/100$ (from bottom to top), $B=1/100$, $M=1$, $q=1$.}\label{figA3}
\end{figure*}

\begin{figure*}
\resizebox{0.7 \linewidth}{!}{\includegraphics*{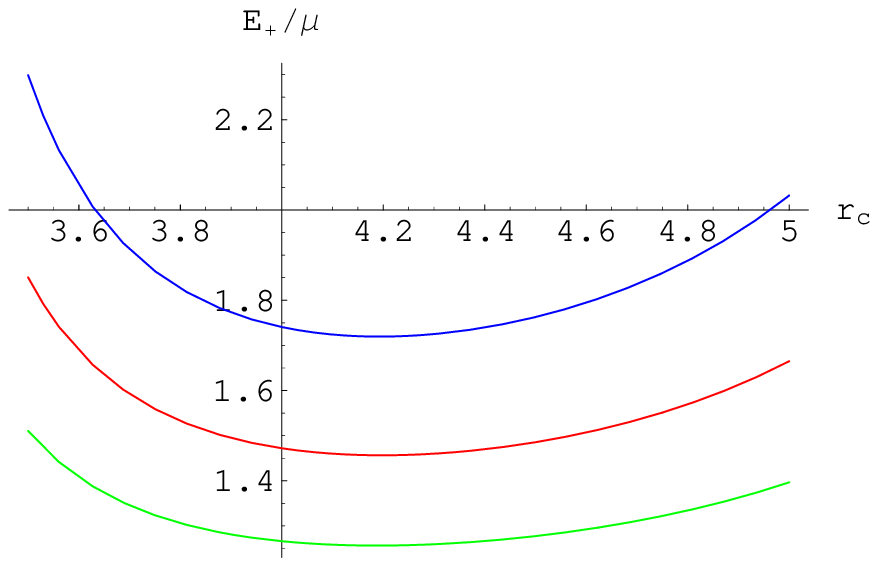}\includegraphics*{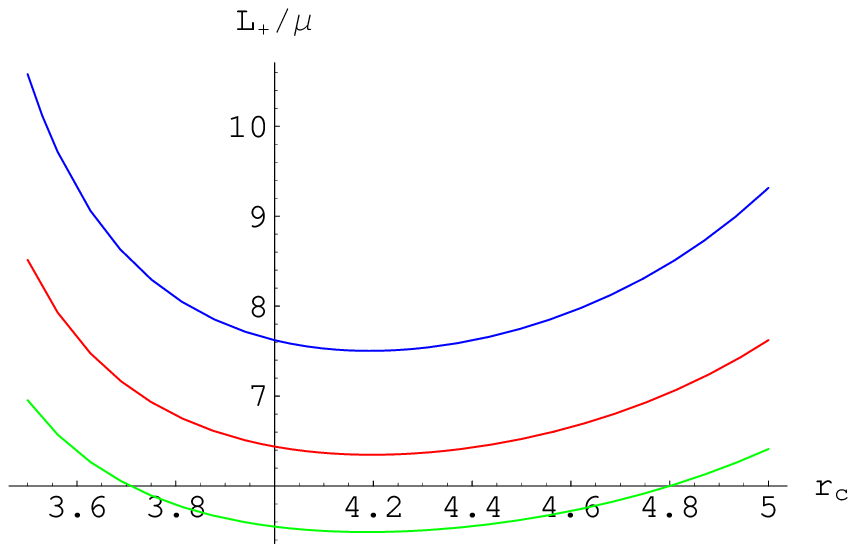}}
\caption{Energy $E_{+}$ (left) and momentum  $L_{+}$ (right) as a function of circular orbit radius for various values of the particle charge $q = -5$(green), $0$(red), $5$ (blue), $\mathcal{E} =1/100$, $B=1/100$, $M=1$.}\label{figA4}
\end{figure*}

For a more general case $B \neq 0$ and $\mathcal{E} \neq 0$ (but still in the approximation of the neglected geometrical influence of the magnetic field onto the metric), the explicit expressions for the energy and momentum are rather cumbersome, therefore
figs. (\ref{figA1}-\ref{figA4}) were used to depict the dependence of energy and momentum on the radius of circular orbits.

In figs. (\ref{figA1}) and (\ref{figA3}) one can see that the energy $E$ and absolute values of the momentum $L$  increase when the tidal force  $\mathcal{E}$ increases. Usually, the effective potential has two extrema: one at the maximum of the potential barrier (corresponding to an unstable orbit and not far from the black hole) and one at the minimum which is further from the black hole (corresponding to a stable orbit). At some minimal values of the angular momentum, these two extrema coincide, which corresponds to the  \emph{innermost stable circular orbit}.  Thus, the minimum of the absolute value of angular momentum $L$ in figs. (\ref{figA1}-\ref{figA4})  is located at the innermost stable circular orbit. As it is shown in figs. (\ref{figA1}) and (\ref{figA3}), \emph{the larger the tidal force $\mathcal{E}$, the closer the innermost stable orbit to the black hole}. This could be explained by the fact that the tidal force $\mathcal{E}$ of the surrounding structure is opposite to the gravitational attraction of the black hole, such that the "effective" gravitational attraction becomes weaker and particles can come closer to the black hole yet still remain in a stable orbit.

The dependence on charge $q$ is more complicated as the situation depends on the direction of the Lorentz force. Thus, the anti-Larmor motion (when the Lorentz force is directed outward from the black hole) corresponds to positive $q$ and negative $L$ or to negative $q$ and positive $L$.
The energy and momentum for various $q$ are shown in figs. (\ref{figA2}) and (\ref{figA4}). The positive and negative angular momenta show opposite dependence of the energy on charge $q$: for negative $L$, lines of energy for negative  $q$ always lay above those for positive $q$, while for positive $L$ the situation is reversed, that is, negative charges correspond to smaller $E$ than positive ones.
The same dependence takes place for the absolute value of the angular momenta. In summary, \emph{the energy $E$ and the absolute value of the angular momentum $\mid L \mid$ grow with the increasing absolute value of charge $\mid q \mid$ for Larmor orbits and decreases for anti-Larmor ones}.
As the magnetic field $B$ is coupled to the charge $q$, the dependence on $B$ is qualitatively similar to the dependence on $q$. As the dependence on the magnetic field for neutral particles was considered earlier in \cite{Konoplya:2006qr}, we focused on the dependence of various quantities on the charge and the tidal force. For the relatively small  $q$ (when $q B$ is still small) and $\mathcal{E}$, considered in figs. (\ref{figA1}) and (\ref{figA4}) and in Tables I and II,
 one cannot tell the difference in the location of the innermost stable Larmor and anti-Larmor orbits. That is why we have only one set of values of $r$ in Tables I and II: the corrections to those values are of the order O($B^4$, $\mathcal{E}^2$, $B^2 \mathcal{E}$). This is certainly not the case for large $q$ for which the location of Larmor and anti-Larmor orbits are quite different. We shall consider the regime of large $q$ in the next subsection.

In order to learn how much energy could be extracted from a particle slowly spiralling toward the center, it is necessary to know what the energy of the particle at the innermost stable circular orbit is.  For this purpose it is convenient to use a function called the \emph{binding energy}, which is the amount of energy released by the particle going from the stable circular orbit located at $r$ to the innermost stable circular orbit $r_{is}$.
Thus, the binding energy (in percent) is
\begin{equation}
\text{Binding Energy}  = 100 \frac{E(r)- E(r_{is})}{E(r)}.
\end{equation}

\begin{figure*}
\resizebox{0.9 \linewidth}{!}{\includegraphics*{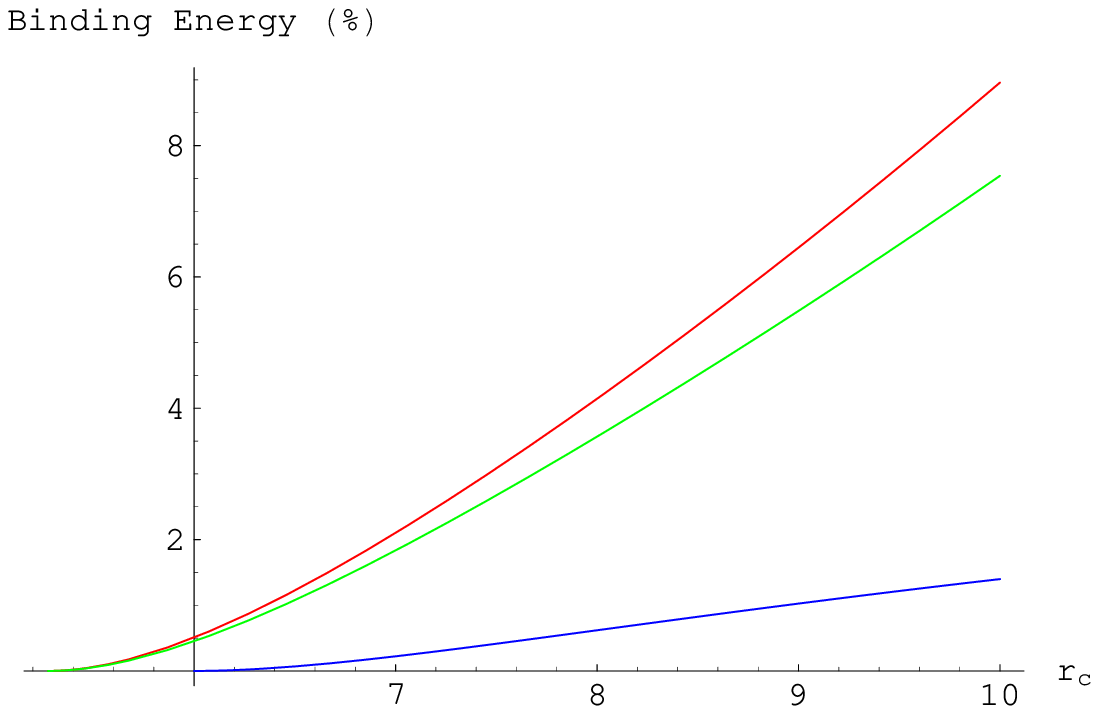}\includegraphics*{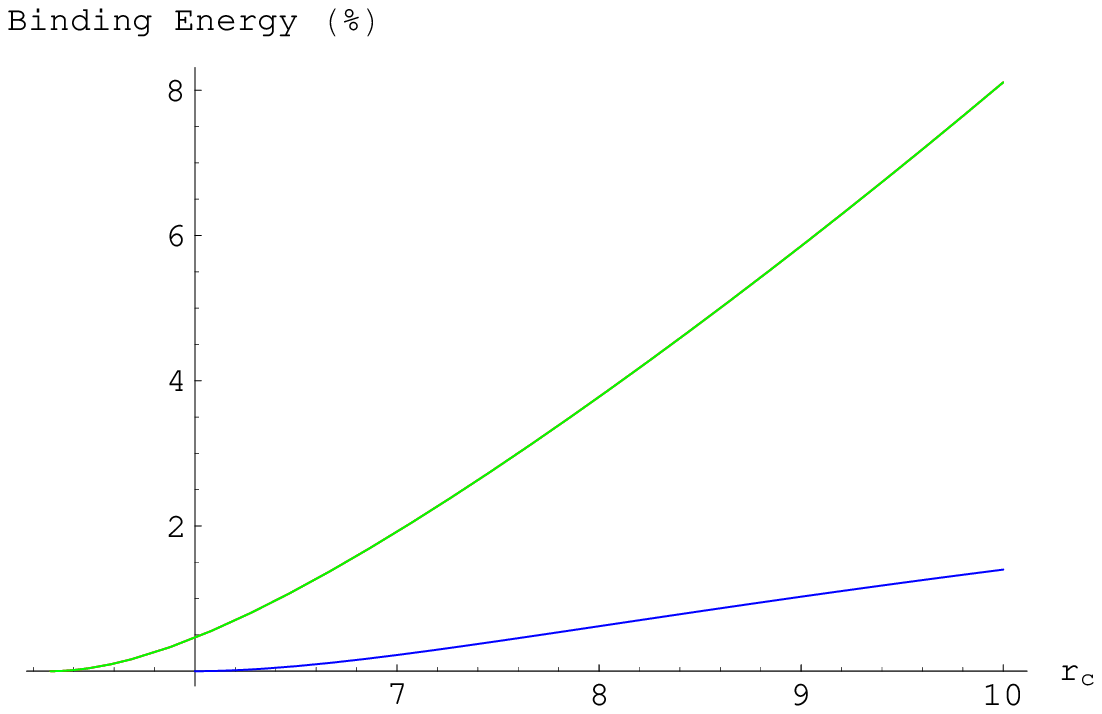}\includegraphics*{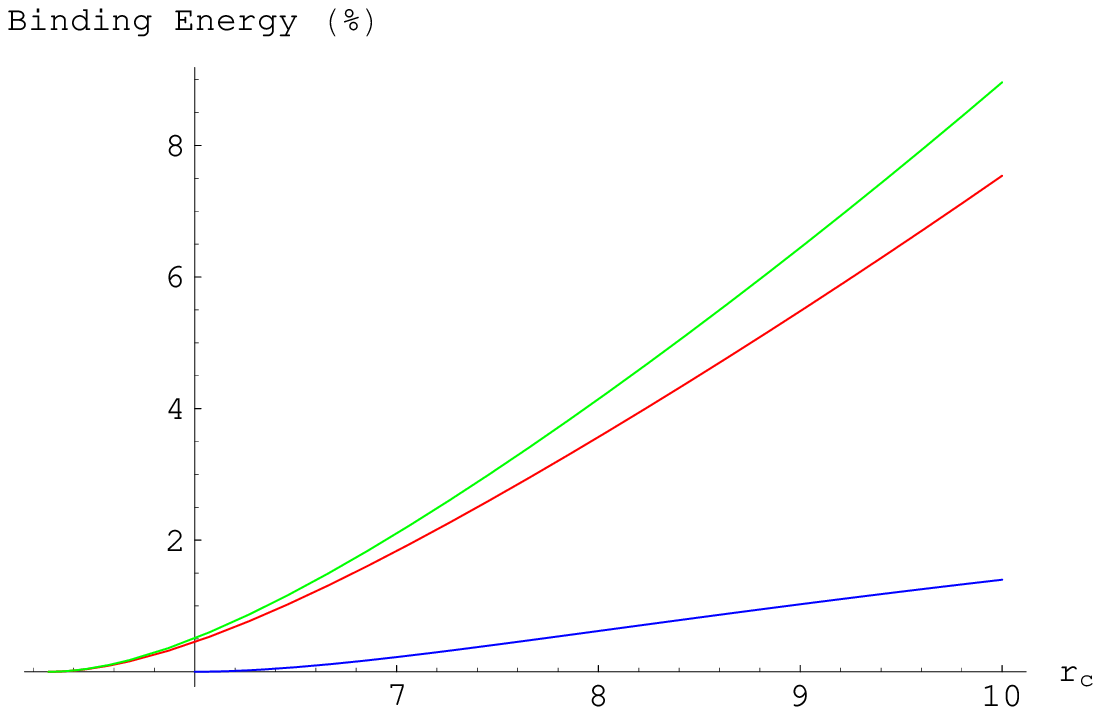}}
\caption{The binding energy as a function of $r$ for $q = -10$ (left), $q=0$ (middle) and $q=10$ (right), $\mathcal{E} = B =10^{-3}$, $M=1$; red is for negative $L$ and green is for positive $L$, blue line is for the Schwarzschild orbit.}\label{figBE1}
\end{figure*}

\begin{figure*}
\resizebox{0.9 \linewidth}{!}{\includegraphics*{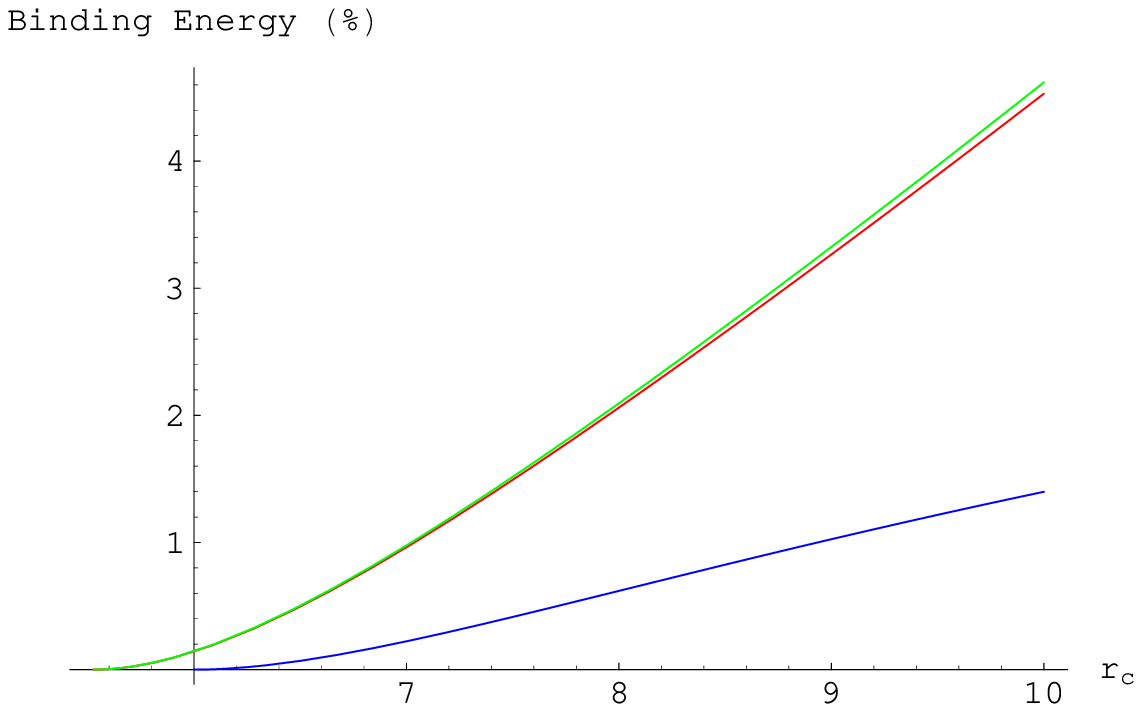}\includegraphics*{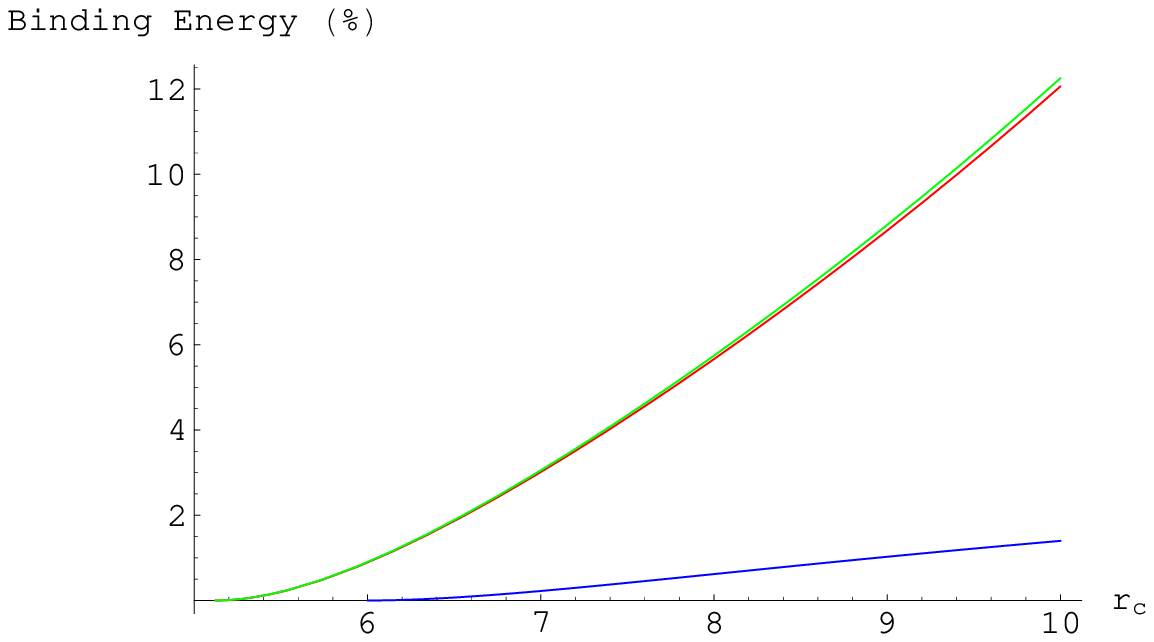}\includegraphics*{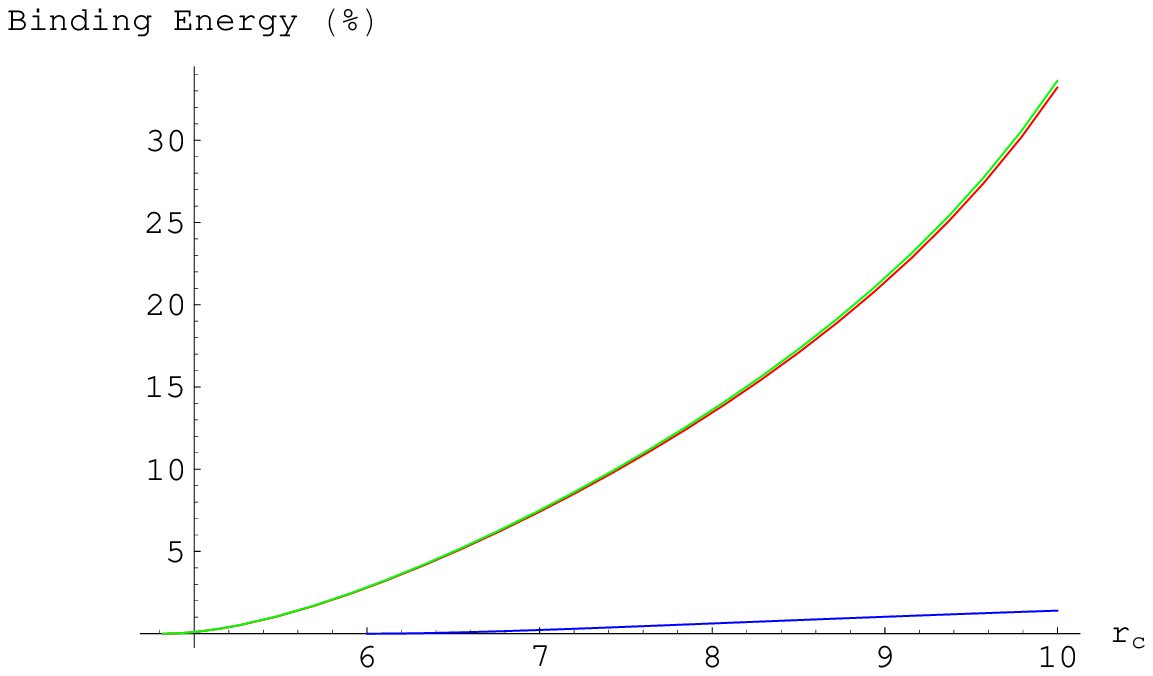}}
\caption{The binding energy as a function of $r$ for $q = 1$, $\mathcal{E}= 0.5 \times 10^{-3}$ (left), $\mathcal{E}=1.5 \times  10^{-3}$ (middle) and $\mathcal{E}=3 \times  10^{-3}$ (right), $B =10^{-3}$, $M=1$;  red is for negative $L$ and green is for positive $L$, blue line is for the Schwarzschild orbit.}\label{figBE2}
\end{figure*}

The binding energy can tell us how much energy the matter (for instance an accretion disk) will release before plunging into the black hole.
For the Schwarzschild black hole (taking $M=1$) the energy on the innermost stable orbit is $E(r_{is}) = \sqrt{8/9}$, so that for a particle coming from
the asymptotic region ($E =1$), the binding energy is about $5.7$ percent. For the Preston-Poisson space-time
there is no asymptotically flat region, so that, the binding energy released at the transition from a given stable orbit $r$ (instead of infinity for the Schwarzschild case) to the innermost stable orbit $r_{is}$ is meaningful.
Figs. (\ref{figBE1}) and (\ref{figBE2})
depict how the binding energy depends on the charge $q$, tidal force $\mathcal{E}$ and magnetic field $B$  as a function of $r$.

In fig. (\ref{figBE1}) the line for negative $L$ lay above the line for positive $L$, if $q$ is negative, and below, if $q$ is positive. Thus, the Larmor orbits have greater binding energy than the corresponding anti-Larmor ones.
Comparison with the blue curve for the Schwarzschild case ($B = \mathcal{E} = 0$) in fig. (\ref{figBE1}) demonstrates  that the magnetic field and its coupling with the charge strongly enhance the release of the binding energy. This property was also observed in \cite{Esteban}
for the black hole in the absence of the tidal force.
The tidal force also strongly increases the binding energy, as can be seen from comparison with the Schwarzschild case in fig. (\ref{figBE2}).
There, the negative momenta correspond to a slightly smaller binding energy than the positive one in fig. (\ref{figBE2}), where the small difference is due to the small $B q$. Thus, the Larmor and anti-Larmor orbits are almost indistinguishable in the regime of tiny $B q$.
In the next subsection we discuss the regime of large charge, when $B q$ is not necessarily small.

\begin{table}
\begin{tabular}{|c|c|c|c|c|c|}
  \hline
  q & r & $L_{+}$ & $L_{-}$ & $E_{-}$ & $E_{+}$ \\
  \hline
  -10 & 5.2745 & 3.64979 & -3.78766 &  0.99508 & 0.95882 \\
  -3 & 5.2841 & 3.69291  & -3.73427 &  0.98163 & 0.97076 \\
  -1 & 5.2849 & 3.70624  & -3.72003 &  0.97792 & 0.97432 \\
  0 & 5.2850 & 3.71308   & -3.71308 &  0.97612 & 0.97612 \\
  1 & 5.2849 & 3.72003   & -3.70624 &  0.97432 & 0.97794\\
  3 & 5.2841 & 3.73427   & -3.69291 &  0.97432 & 0.97794 \\
  10 & 5.2753 & 3.78766  & -3.64979 &  0.95882 &  0.99508 \\
  \hline
\end{tabular}
\caption{Energies $E_{+}$ and $E_{-}$ and the angular momenta $L_{+}$ and  $L_{-}$  at the innermost stable circular orbit
for various values of charge $q$; $B=\mathcal{E}=10^{-3}$, $M=1$}
\end{table}

\begin{table}
\begin{tabular}{|c|c|c|c|c|c|}
  \hline
 $\mathcal{E}/10^{-3}$ & r&  $L_{+}$ & $L_{-}$ &  $E_{-}$ & $E_{+}$ \\
  \hline
  0 & 5.9997 & 3.47018 & -3.45818  & 0.94119 & 0.94445 \\
  $0.5 $ & 5.5277 & 3.60380 & -3.59086  & 0.95811 & 0.96156 \\
  $1$ & 5.2849 & 3.72003 & -3.70624  & 0.97432 & 0.97794 \\
  $1.5$  & 5.1213 & 3.82919 & -3.81458  & 0.99046 & 0.99425 \\
  $2$  & 4.9982 &  3.93531 & -3.91988 & 1.00684 & 1.01080 \\
  $2.5$  & 4.8994 & 4.04064 &  -4.02438 & 1.02366 & 1.02779 \\
  $3$  & 4.8168 & 4.14672 & -4.1296  & 1.04105 &  1.04537 \\
  \hline
\end{tabular}
\caption{Energies $E_{+}$ and $E_{-}$ and the angular momenta $L_{+}$ and  $L_{-}$ at the innermost stable circular orbit for various values of charge $\mathcal{E}$; $B=10^{-3}$, $q=1$, $M=1$}
\end{table}

\subsection{The general case with all non-vanishing parameters}

When the charge $q$ is large enough, so that the term $B q$ is no longer small, one cannot safely neglect the geometric influence of the magnetic field, because even subdominant terms containing $B q$ can be quite large in comparison, for instance, with the tidal force corrections $\sim \mathcal{E}$. Thus, we considered the full metric (\ref{PPmetric}) with a nonzero magnetic field both in the metric coefficients and in the Hamilton-Jacobi equation for large $q$. As shown in Table III, the locations of Larmor and anti-Larmor orbits already differ a lot
for $B q \sim 1$. At the same time, the data for $B q \sim 10^{-2}$ (the first line in the Table III) still almost coincide with
the approximation of the neglected "geometric contribution" of $B$ considered above. Another distinction from
the above small charge approximation is that, while for small  $B q$ the energy at the innermost stable orbit is only slightly smaller for the anti-Larmor orbits than for the Larmor ones (Tables I and II), for moderate $B q$, this difference is very large (see last line in Table III) and the Larmor orbits may have one order larger energy and momentum than the anti-Larmor ones.

In the case of large $q$, the binding energy of Larmor orbits is not always greater than the anti-Larmor ones as it depends on a stable orbit from which the particle plunges into the black hole. For more distant stable orbits (see fig. (\ref{figBE2})) the binding energy of the anti-Larmor orbits can be larger than the corresponding Larmor ones.
If one remembers that the Larmor motion is  a cyclotron rotation in a uniform magnetic field perturbed by a black hole, then, it becomes clear that in the case of large $B q$ the "perturbation" by the black hole is relatively small, so that the rotation is almost purely cyclotronic, especially at a relatively large distance from the black hole.

\begin{table}
\begin{tabular}{|c|c|c|c|c|c|c|}
  \hline
  q & $r_{+}$ & $r_{-}$ & $L_{+}$ & $L_{-}$ & $E_{-}$ & $E_{+}$ \\
  \hline
  -10 & 5.27452 & 5.27527  & 3.64981 & -3.78768 & 0.995088  & 0.95882 \\
  -50 & 5.04704 &  5.11032  & 3.49905  & -4.19065 & 1.08802  & 0.90389 \\
  -100 & 4.86491 & 4.61837  & 3.46307  & -4.87639 & 1.23937  & 0.85679 \\
  -1000 & 2.74943 & 4.32074   & 5.20535   & -24.8736 &  5.90001 & 0.59092  \\
  \hline
\end{tabular}
\caption{Energies $E_{+}$ and $E_{-}$ and the angular momenta $L_{+}$ and  $L_{-}$  at the innermost stable circular orbit
for various large values of charge $q$; $B=\mathcal{E}=10^{-3}$, $M=1$}
\end{table}

\begin{figure}
\resizebox{0.8 \linewidth}{!}{\includegraphics*{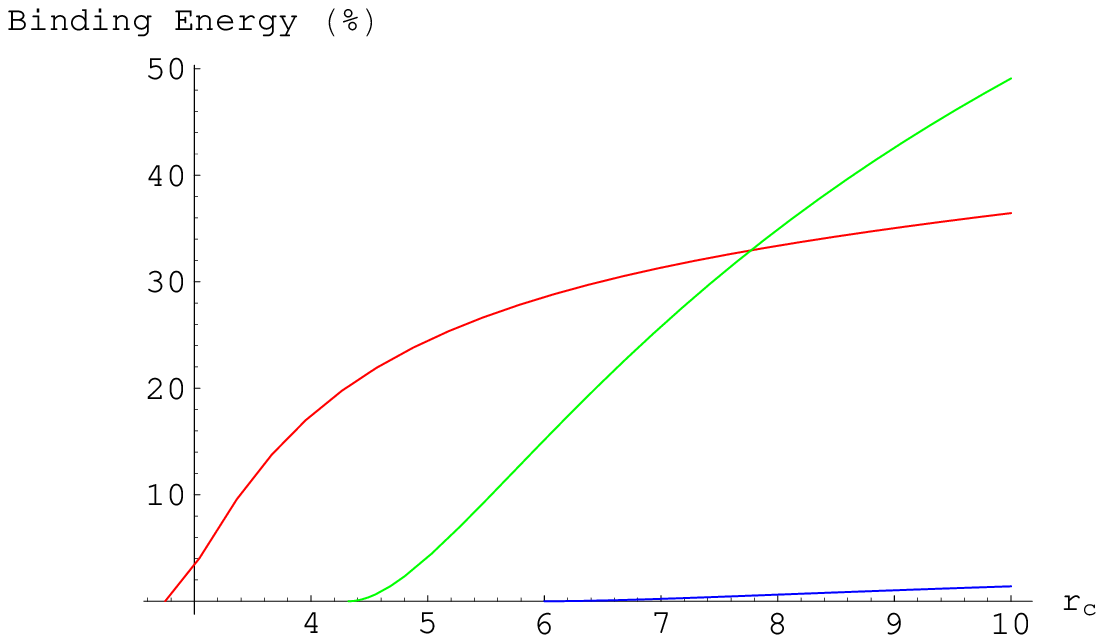}}
\caption{The binding energy as a function of $r$ for $q = -1000$, $\mathcal{E} = B =10^{-3}$, $M=1$, red line corresponds to the orbit with a negative momentum (Larmor for $q <0$), green corresponds to a positive (anti-Larmor)}\label{figBE3}
\end{figure}

Finally, let us discuss the radial stability of the considered circular orbits. The condition of the radial stability can be written
as:
\begin{equation}
\frac{\partial^2 U_{eff}}{\partial r^2} > 0 \quad \text{- stability condition.}
\end{equation}
From numerical study of the above inequality (when $U_{eff}$ is given by its most general expression (\ref{general-eff-pot})), we conclude that at sufficiently large $|q| B$, the region of stability of anti-Larmor orbits
can approach the event horizon,
\begin{equation}
r > r_{h} \quad \text{- anti-Larmor, high $|q| B$,}
\end{equation}
while Larmor orbits are only stable for $r \gtrapprox 4. 2 M$ for zero tidal force $\mathcal{E}=0$, approaching $r \gtrapprox 3.6 M$ at $\mathcal{E}=1/50 M$,
\begin{equation}\label{38}
r \gtrapprox  \left(\begin{array}{c}
                                4.2 M \text{($\mathcal{E}=0$)} \\
                                3.6 M \text{($\mathcal{E}=1/50 M$)}\\
                              \end{array}
                            \right)
\text{- Larmor, high $|q| B$.}
\end{equation}

This is in good agreement with the  $r \gtrapprox 4. 3 M$ limit observed in \cite{Galtsov-book} for the Ernst solution in the absence of the tidal force. The small difference between (\ref{38}) and \cite{Galtsov-book} is apparently due to the neglected
contribution of orders $B^3$  and higher in the Preston-Poisson metric.
The new feature here is that the tidal force can considerably expand the region of stability of the Larmor orbits making them closer to the black hole horizon. This can also be seen from the Table II.

\section{Scalar field perturbations in the Preston-Poisson space-time}

The aim of our study of a test scalar field in the Preston-Poisson space-time is to find the proper oscillation frequencies (quasinormal modes) which dominate in the response to the perturbation at late time.
As shown in \cite{Konoplya:2006rv}, the quasinormal modes of black holes, when considered in the astrophysical context, are essentially independent of the behavior of the master wave equation far from the black hole. The explanation is straightforward: quasinormal modes are poles of the reflection coefficient of the scattering process which occurs near the peak of the effective potential, so that the low-laying, dominating modes are "localized" near the maximum of the effective potential. For the Schwarzschild solution this happens at $r = 3 M$, while for the Preston-Poisson space-time this value is only slightly corrected. The wave equation does not have physical meaning at $r=\infty$, because for astrophysical processes of this kind, "infinity" effectively is situated at the distance  which is much larger than the radius of the black hole. In our case, "infinity" is at least quite a few times larger than the black hole radius and still far from the edge of the torus, $r_h \ll r \ll a$. We have considered the same spatial region when analyzing particle's motion.

For a test neutral scalar field, influence of the magnetic field $B$ can usually be neglected, because the energy density of the magnetic field is much smaller than that of the gravitational one.
When $B=0$, the coefficients of the Preston-Poissin metric (\ref{PPmetric}) are simplified as follows:
\begin{align}\label{PPcoef-simpl}
g_{vv}&=-f+\mathcal{E}(r-2M)^2(3\cos^2-1),\\
g_{vr}&=1,\\
g_{v\theta}&=-2\mathcal{E} r^2(r-2M)\sin\theta\cos\theta,\\
g_{\theta\theta}&=r^2+\mathcal{E} r^2 (r^2-2M^2)\sin^2\theta,\\
g_{\varphi\varphi}&=r^2\sin^2\theta-\mathcal{E} r^2(r^2-2M^2)\sin^4\theta. \label{PPcoef-simpl1}
\end{align}

The Klein-Gordon equation in a curved space-time has the following form
\begin{equation*}
\square\Psi= \frac{\partial}{\partial x^{\mu}}\left(g^{\mu\nu}\sqrt{-g}\frac{\partial \Psi}{\partial x^{\nu}}\right)=0.
\end{equation*}

\begin{figure*}
\resizebox{0.7 \linewidth}{!}{\includegraphics*{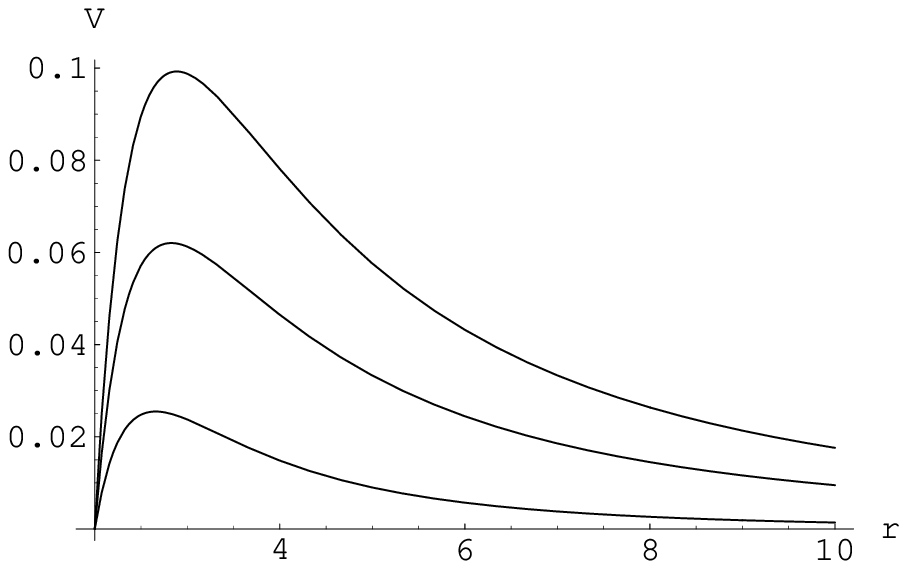}\includegraphics*{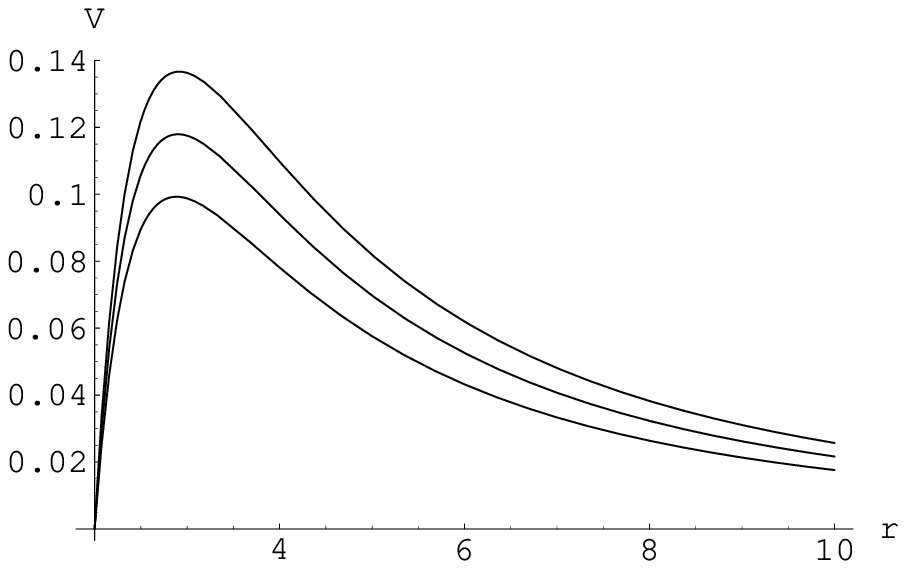}}
\caption{Effective potentials for a) $\ell =1$, $m=0$, $\mathcal{E} = 0, 1/10, 2/10$ (from the top to bottom), b)
$\ell =1$, $m=1$, $\mathcal{E} = 0, 1/10, 2/10$ (from bottom to top). The potential barrier is raised by the tidal force for non-zero $m = \pm 1$ and is lowered at the lowest modes ($m=0$).}\label{fig1}
\end{figure*}

In the general case, the variables in the Klein-Gordon equation cannot be decoupled for the above metric.
However, a further assumption can remedy the situation. We shall assume, that as the torus is situated
far from the black hole $M \ll a$, the tidal force $\mathcal{E}$, being relatively small, acts almost homogeneously in
a small region where the low laying quasinormal modes are "localized", that is near the peak of the effective potential
\begin{equation}\label{rmax}
r = r_{max} = p M, \quad p =\frac{8}{3}, \quad \ell = 0,
\end{equation}
$$ p = \frac{3}{2} + \frac{\sqrt{9 + \ell (\ell +1) (14 + 9 \ell (\ell +1))}- 3}{2 \ell (\ell +1)}, \quad \ell \neq 0.$$
In other words, the tidal force at the potential peak does not change much when one slightly moves
away from the peak $r \approx r_{max}$.
 Thus, we assume that $M \ll a$ holds "with a margin", being also  $r_{max} \ll a$. In this case, being interested only in the region near the peak, we can expand \emph{the perturbed part} of the Preston-Poisson metric (\ref{PPcoef-simpl}-\ref{PPcoef-simpl1}) (i.e. the terms containing $\mathcal{E}$) in powers of ($\mathcal{E}, (r- r_{max})$).
Further we neglect the  corrections to the Schwarzschild metric of order $\mathcal{E} (r - r_{max})$ and higher,
which means that, as we assumed, the tidal force barely changes near the peak.
Although such an approximation does not allow us to make accurate calculations of quasinormal spectrum, we can still
estimate the dominating modes numerically. In reality the tidal force is smaller on the left (at the black hole side) of the peak
and larger on the right of it, so that the used expansion in powers of  $(\mathcal{E}, (r - r_{max}))$ is a kind of "averaging" of
the tidal force in the small region of the localization of the dominant modes.

In the above approximation the metric can be reduced to the diagonal form
\begin{equation}\label{diagonal1}
ds^2=g_{vv}dt^2-\frac{1}{g_{vv}}dr^2+g_{\theta\theta}d\theta^2+g_{\varphi\varphi}d\varphi^2,
\end{equation}
through the following transformations,
\begin{equation}
v = t + F(r), \quad F'(r) g_{vv}(r, \theta) + 1 =0,
\end{equation}
where
\begin{equation}
g_{vv}=-f,\qquad \qquad g_{v\theta}=0.
\label{equation_order_correction}
\end{equation}
The determinant of the diagonal metric (\ref{diagonal1}) is
\begin{equation*}
\sqrt{-g}=\sqrt{g_{\theta\theta}g_{\varphi\varphi}}\approx r^2\sin \theta+O(\mathcal{E}^2).
\end{equation*}

Due to the Killing vectors in $t$ and $\phi$, the stationary ansatz for the perturbation is implied:
\begin{equation*}
\Psi(t,r,\theta,\varphi)=e^{im\varphi-i\omega t} {\widetilde{\Phi}}(r,\theta).
\end{equation*}
The wave equation can be written as follows
\begin{align*}
\omega^2 \left(1-\frac{2M}{r}\right)^{-1}r^2 \widetilde{\Phi}+\frac{\partial}{\partial r}\left(\left(1-\frac{2M}{r}\right)r^2 \frac{\partial\widetilde{\Phi}}{\partial r}\right)+ S \widetilde{\Phi}=0,
\end{align*}
where
\begin{align*}
S\widetilde{\Phi}=\frac{1}{\sin\theta} \left( \frac{\partial}{\partial\theta}\left(\sqrt{\frac{g_{\varphi\varphi}}{g_{\theta\theta}}}\frac{\partial\widetilde{\Phi}}{\partial\theta}\right)
       +(-m^2)\widetilde{\Phi}\sqrt{\frac{g_{\theta\theta}}{g_{\varphi\varphi}}} \right).
\end{align*}

Expansion of $\sqrt{\frac{g_{\theta\theta}}{g_{\varphi\varphi}}}$ and $\sqrt{\frac{g_{\varphi\varphi}}{g_{\theta\theta}}}$ in powers of ($\mathcal{E},(r-r_{max})$) yields
\begin{align*}
\sqrt{\frac{g_{\varphi\varphi}}{g_{\theta\theta}}}&=\sin\theta- (p^2 -2)\mathcal{E} M^2 \sin^3\theta+O(\mathcal{E}(r-r_{max}),\mathcal{E}^2),\\
\sqrt{\frac{g_{\theta\theta}}{g_{\varphi\varphi}}}&=\frac{1}{\sin\theta}+(p^2 -2) M^2\mathcal{E} \sin\theta +O(\mathcal{E}(r-r_{max}),\mathcal{E}^2).
\end{align*}

Then, the angular part can be rewritten as
$$\frac{1}{\sin\theta}\frac{\partial}{\partial\theta}\left(\sin\theta\frac{\partial\widetilde{\Phi}}{\partial\theta}\right)-\frac{(p^2 -2)\mathcal{E} M^2}{\sin\theta}\frac{\partial}{\partial \theta}\left(\sin^3\theta\frac{\partial\widetilde{\Phi}}{\partial\theta}\right) $$
\begin{equation}\label{angular}
-m^2 \left(\frac{1}{\sin^2\theta}+(p^2 -2)\mathcal{E} M^2\right)\widetilde{\Phi}=S\widetilde{\Phi}.
\end{equation}

Introducing the ``tortoise'' coordinate $r^*$, which is defined by the relation $dr^*=\frac{dr}{f(r)}$, together with a new wave function $\widetilde{\Phi}=\frac{\Phi}{r}$, one can reduce the wave equation to the Schrodinger wavelike form with an effective potential $V(r, S)$,
\begin{equation}\label{wavelike}
\frac{d^2\Phi}{{dr^*}^2}+\left(\omega^2-V(r, S)\right)\Phi=0,
\end{equation}
where
\begin{equation}\label{potential}
V(r,S)= \left(1-\frac{2 M}{r}\right)\left(\frac{S}{r^2} - \frac{2M}{r^3} \right).
\end{equation}

The latter effective potential is identical to that of the Schwarzschild, up to the different
angular eigenvalues $S$, which will be calculated in the next section.

\section{Angular wave equation}

The eigenvalues of the angular equation (\ref{angular}) can be calculated via two methods: the convergent and accurate Forbenius method and the
method of expansion into the Associated Legendre Polynomials. Here we shall imply that the maximum of the effective potential is situated at its Schwarzschild distance $r_{max}$ given by (\ref{rmax}) for each value of the multipole number $\ell$, i.e. we shall neglect tiny displacement of the peak induced by the tidal force $\mathcal{E}$. This approximation is fully justified as the inclusion of ($\mathcal{E}$, $B$) corrections to the maximum's location contributes only at the subdominant orders, as discussed in Sec. VII.

\subsection{Frobenius Method}

Equation (\ref{angular}) is reduced to the Heun's equation by setting $x=\sin^2\theta$ and thus giving
\begin{equation}
\frac{d^2 \widetilde{\Phi}}{dx^2}+\frac{2-3 x-4 x \sigma +5 x^2 \sigma }{2 (x-1) x (\sigma x-1 )}\frac{\widetilde{\Phi}}{dx}-\frac{m^2 + S x+m^2 x \sigma }{4 (x-1) x^2 (\sigma x-1 )} \widetilde{\Phi}=0,
\label{Equation_wave_equation_angular_x_sine}
\end{equation}
where
$$\sigma= (p^2 -2) \mathcal{E}  M^2.$$
The above expression for $\sigma$ will be considered as a constant in the further calculations, which is justified by the approximation (\ref{rmax}).   In this way, we are only able to find even eigenfunctions for $\theta \rightarrow -\theta$. The odd eigenfunctions must be found  separately.
Setting $\widetilde{\Phi}=x^{\frac{m}{2}}\psi(x)$, one can transform equation (\ref{Equation_wave_equation_angular_x_sine}) into the Heun's equation
$$ \frac{d^2 \psi}{dx^2}+\left(\frac{\gamma}{x}+\frac{\delta}{x-1}+\frac{\kappa}{x-a_H}\right)\frac{d \psi}{dx}+ $$

\begin{equation}
\frac{\alpha\beta x-q}{x(x-1)\left(x-a_H\right)}\psi=0,
\end{equation}
with
\begin{align*}
\gamma&=m+1,\qquad\delta=\frac{1}{2},\qquad\kappa=1,\qquad \\
q&=\frac{m^2+m+\tilde{S}+2m^2\sigma+2m\sigma}{4\sigma}, \\
 \alpha&=\frac{m}{2},\qquad\beta=\frac{m+3}{2},\qquad a_H=\frac{1}{\sigma},
\end{align*}
and
\begin{equation*}
\omega=\gamma+\delta-1=\alpha+\beta-\kappa=m+\frac{1}{2}.
\end{equation*}
The solution to the Heun's equation can be expanded as follows
\begin{equation*}
\psi(x)=\sum _{n=-\infty }^{+\infty }c_n y_n(x),
\end{equation*}
if $c_n$ satisfies the recurrence relations
\begin{equation*}
F_n c_{n-1}+G_n c_n+H_n c_{n+1}=0,
\end{equation*}
where
\begin{widetext}
\begin{align*}
F_n&=\frac{(n+\alpha-1)(n+\beta-1)(n+\gamma-1)(n+\omega -1)}{(2n+\omega -2)(2n+\omega -1)},\\
G_n&=-\frac{\kappa n(n+\omega )(\gamma-\delta)+(n(n+\omega )+\alpha\beta)(2n(n+\omega )+\gamma(\omega -1))}{(2n+\omega +1)(2n+\omega -1)}+n(n+\omega)a_H+q,\\
H_n&=\frac{(n+1)(n+\omega -\alpha+1)(n+\omega -\beta+1)(n+\delta)}{(2n+\omega +2)(2n+\omega +1)},
\end{align*}
\end{widetext}
with $n=\frac{\ell-m}{2}$.\\
Since the coefficients of $c_{n-1}$, $c_n$ and $c_{n+1}$ vanish independently, the eigenvalue can be obtained by taking $G_n=0$,
\begin{equation}\label{S-formula}
S=\textbf{-}\frac{\ell (1+\ell) \left(3-4 \ell-4 \ell^2+2 \ell \sigma +2 \ell^2 \sigma -6 m^2 \sigma \right)}{(-1+2 \ell) (3+2 \ell)}.
\end{equation}

The above formula has been derived here for the "even" modes, i.e. modes with even values of $\ell - |m|$, though, using the expansion
in terms of the Associated Legendre Polynomial, we show that this formula is valid also for odd eigenfunctions.

\subsection{Expansion in terms of the Associated Legendre Polynomial}

Introducing a new variable $x$ in a different way,  $x=\cos\theta$, equation (\ref{angular}) can be written as
$$ \frac{d^2\widetilde{\Phi}}{dx^2}+\frac{2 x \left(2 x^2 \sigma -2 \sigma +1\right)}{(x-1) (x+1) \left(x^2 \sigma -\sigma +1\right)}\frac{d\widetilde{\Phi}}{dx} $$
\begin{equation}\label{54}
+\frac{-m^2 x^2 \sigma +  S x^2-m^2 \sigma -m^2-S}{(x-1)^2 (x+1)^2 \left(x^2 \sigma -\sigma +1\right)} \widetilde{\Phi}=0.
\end{equation}
Since the Preston-Poisson solution is accurate through the order of $\mathcal{E}$, one can expand the angular eigenfunction function and the corresponding eigenvalues in terms of $\sigma = (p^2 -2) \mathcal{E} M^2$ as
\begin{align}
\widetilde{\Phi}&=P^{\ell}_{m}
+\sigma\left(\sum_{\ell\geq|m|}C^{\ell}_{m}P^{\ell}_{m}\right)+O[\sigma^2],
\label{Equation_expansion_wave_equation}\\
S&=\ell(\ell+1)+\sigma S',
\label{Equation_expansion_eigenvalue}
\end{align}
where $\ell =0, 1, 2, .....$ and $m = - \ell, -\ell +1,.....,\ell -1, \ell$.

Substituting equations (\ref{Equation_expansion_wave_equation}) and (\ref{Equation_expansion_eigenvalue}) into the wave equation (\ref{54}), one can obtain the $S'$ values, which are summarized for lower $\ell$ and $m$
in Table \ref{Table_Eigenvalue}.
\begin{table}[h]
\centering
    \begin{center}
        \begin{tabular}{|c|c|ccc|ccccc|}
        \hline
        {\scriptsize$\ell=$}&{\scriptsize$0$}&{\scriptsize$ $}&{\scriptsize$1$}&{\scriptsize$ $}&{\scriptsize$ $}&{\scriptsize$ $}&{\scriptsize$2$}&{\scriptsize$ $}&{\scriptsize$ $}\\
        \hline
        {\scriptsize$m=$}&{\scriptsize$0$}&{\scriptsize$-1$}&{\scriptsize$0$}&{\scriptsize$1$}&{\scriptsize$-2$}&{\scriptsize$ -1$}&{\scriptsize$0$}&{\scriptsize$1$}&{\scriptsize$2$}\\
        \hline
        {\scriptsize$S'=$}&{\scriptsize$0$}&{\scriptsize$\frac{4}{5}$}&{\scriptsize$-\frac{8}{5}$}&{\scriptsize$\frac{4}{5}$}&{\scriptsize$\frac{24}{7}$}&{\scriptsize$-\frac{12}{7}$}&{\scriptsize$-\frac{24}{7}$}&{\scriptsize$-\frac{12}{7}$}&{\scriptsize$\frac{24}{7}$}\\
        \hline
        \end{tabular}
    \end{center}
\caption{The eigenvalues of $\ell=0$,$1$ and $2$.}
\label{Table_Eigenvalue}
\end{table}

Table \ref{Table_Eigenvalue} shows that the formula (\ref{S-formula}) is indeed valid for the low laying even and odd modes.
With this data for the angular eigenvalues at hand, we can start numerical calculations of the quasinormal modes.

In principle, one can have a slightly better approximation of the estimation of the angular eigenvalue $S$, if the tiny shifts of the value of the radial coordinate $r$ in the maximum of the effective potential due
to nonzero values of $\mathcal{E}$ are taken
into account, when calculating the tidal force in the potential maximum. This means that instead of Eq. (\ref{rmax}) for $p$, one could use a general solution for the maximum of the effective potential (\ref{potential}):
\begin{equation}\label{p-full}
p = -3 + 3 S + \sqrt{(9 + S (14 + 9 S))/2 S}
\end{equation}
Comparison of the QNMs computed with the Schwarzschild values of $p$ and with (\ref{p-full}) for lower values of $\ell$,
reveal a very small difference of about $0.01$ percent, which cannot be distinguished in figs. (\ref{fig2}) and (\ref{fig3}).
In the next section, the more accurate value given in (\ref{p-full}) shall be used.

\section{Estimations of the quasinormal modes}

An essential moment in the determination of the quasinormal modes is the agreed notion of asymptotical region, "infinity", which is simply a distance
much larger than the black hole radius. Once the near horizon and far regions are defined, we can formulate the boundary conditions for quasinormal modes. Quasinormal modes are solutions to the wavelike equation (\ref{wavelike}) which are purely ingoing waves at the event horizon and purely outgoing waves at the far asymptotic region:
$$
\Phi \sim e^{- i \omega r^{*}}, \quad r^{*} \rightarrow - \infty, \quad
$$
\begin{equation}
\Phi \sim e^{+ i \omega r^{*}}, \quad r^{*} \rightarrow \quad far \quad asymptotic \quad region
\end{equation}

Thereby, no incoming waves are allowed either from the horizon or from the far region. This means that quasinormal modes correspond to the proper oscillations of the black hole's response to the external perturbation at late time, that is, when the source of the initial perturbation does not act anymore. In other words, the perturbation is considered as a "momentary".

\begin{figure*}
\resizebox{0.7 \linewidth}{!}{\includegraphics*{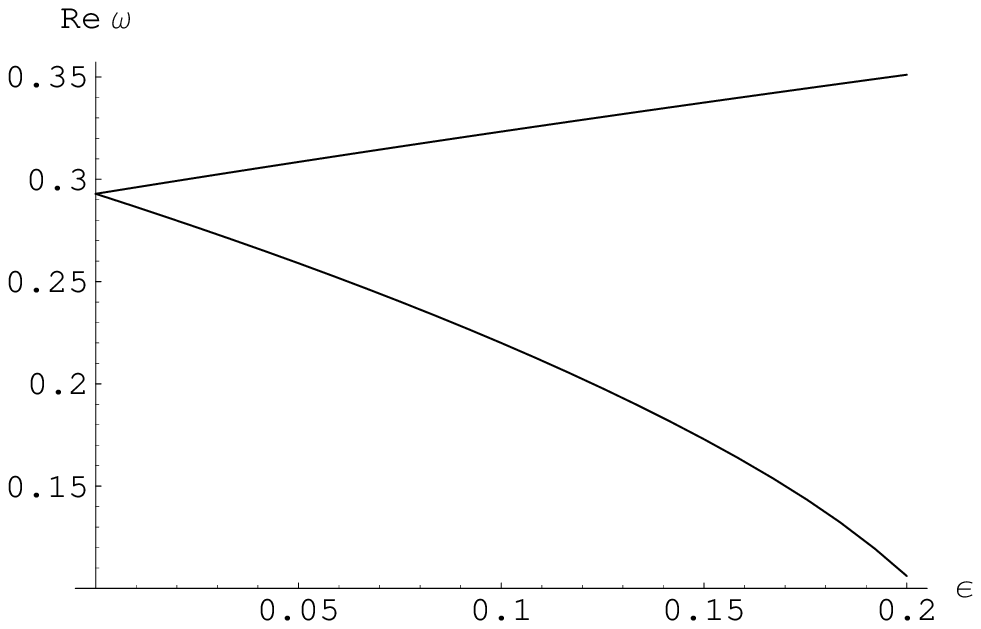}\includegraphics*{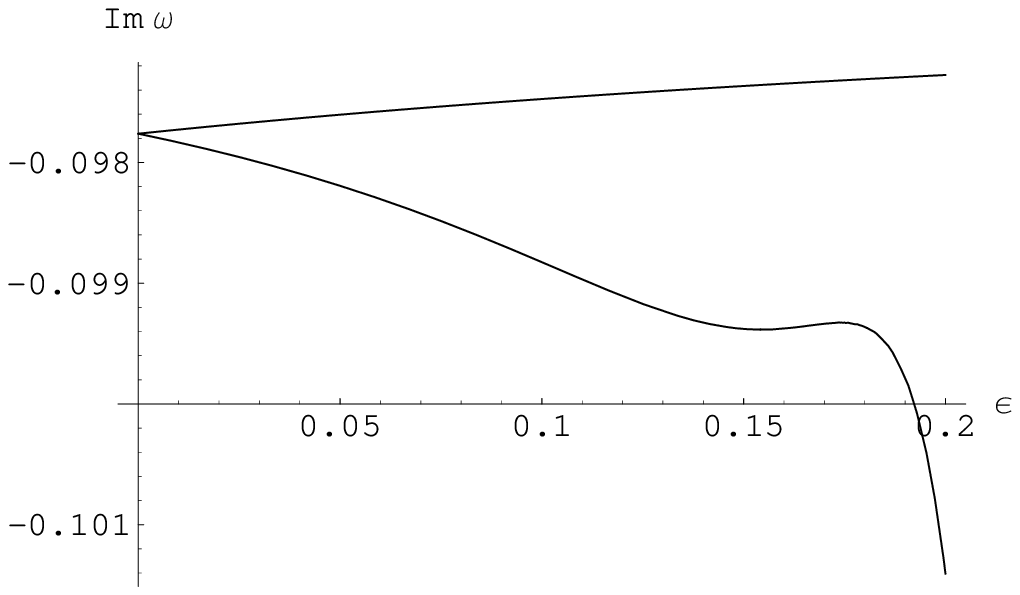}}
\caption{Real (left) and imaginary (right) parts of QNMs for $\ell =1$, $m=0$ (bottom) and
$\ell =1$, $m=1$ (top).}\label{fig2}
\end{figure*}

\begin{figure*}
\resizebox{0.7 \linewidth}{!}{\includegraphics*{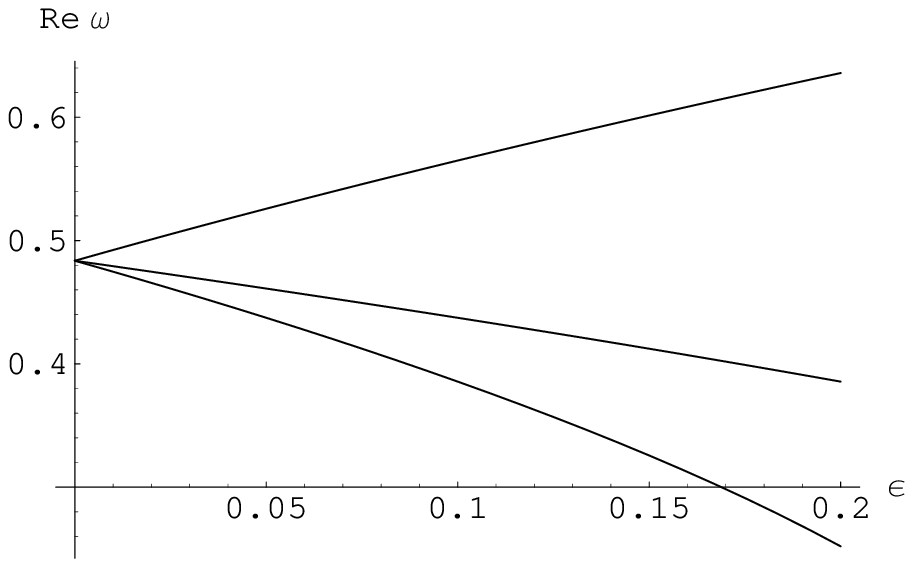}\includegraphics*{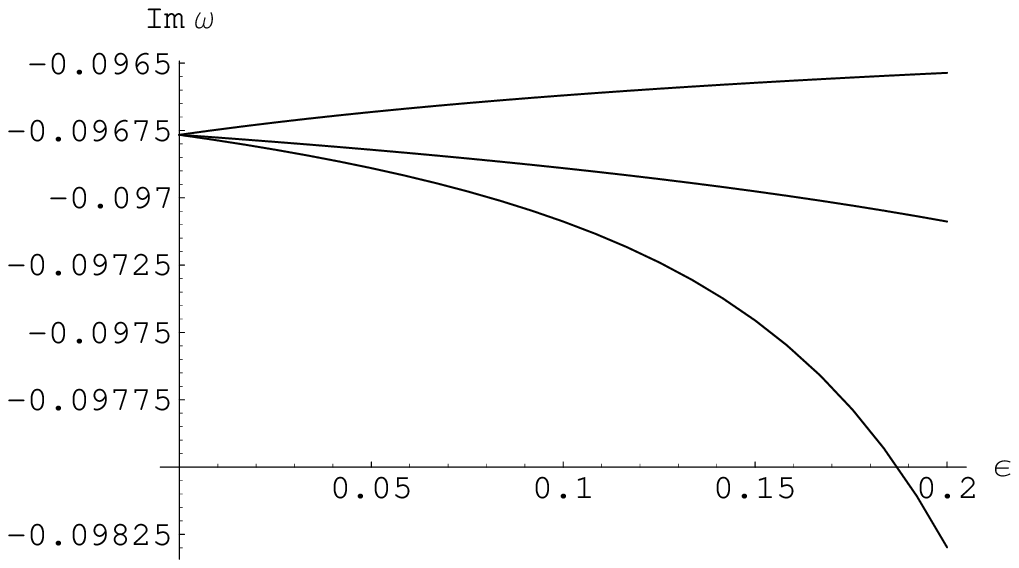}}
\caption{a. Left panel: Real part of $\omega$ for $\ell =2$, $m=0$ (bottom),
$m=1$, and $m=2$ (top), b. Right panel: Imaginary part of $\omega$ for $\ell =2$, $m=0$ (bottom),
$m=1$, and $m=2$ (top)}\label{fig3}
\end{figure*}

In order to calculate quasinormal modes defined above, one can use the WKB formula,
which takes into consideration the behavior of the perturbation near the top of the effective potential.
The asymptotic region is located at a distance, which is at least one order larger than the black hole radius,
such that one could qualitatively model the processes far from the black hole. At the same time, such a defined "infinity" must be
located at a distance which is quite a few times smaller than the radius of the giant torus. The latter is necessary to avoid
influence of the boundary effects near the torus of matter. Thus, there should be a "valley" far from the black hole and far from the edge of the torus, $r_{h} \ll r \ll a $ which could model an asymptotic region.

The QN spectrum is one of the essential
characteristics of a black hole because it depends only on the black hole
parameters and not on the way by which modes were excited.  Supposing
$\Psi \sim e^{-i \omega t}$, quasinormal modes can be written in the form
$$
	\omega = \omega_{Re} - \omega_{Im} i,
$$
 where positive  $\omega_{Im}$ is proportional to the decay rate of a damped
 QN mode.  The low laying QN frequencies have the smallest decay rates in
the spectrum and thus dominate in a signal at a sufficiently late time.  They
can be calculated by the WKB approach \cite{WKB}, \cite{WKBorder}.
Introducing $Q = \omega^2 - V$, the 6-th order WKB formula reads
\begin{equation}\label{WKBformula}
	\frac{i Q_{0}}{\sqrt{2 Q_{0}''}} - \sum_{i=2}^{i=6}
		\Lambda_{i} = n+\frac{1}{2},\qquad n=0,1,2\ldots,
\end{equation}
 where the correction terms $\Lambda_{i}$ were obtained in \cite{WKB},
 \cite{WKBorder}.  Here $Q_{0}^{i}$ means the i-th derivative of $Q$ at its
 maximum with respect to the tortoise coordinate $r^\star$, and $n$ labels
 the overtones. The WKB formula (\ref{WKBformula}) was effectively used in a
 lot of papers (see \cite{WKBuse} and references therein).

As an additional check of the WKB data, we used the time domain integration scheme, which was related, for instance, in \cite{timedomain} and used in a number of other works (see  \cite{time-domain-apply} and references therein).
The obtained time domain data shows about a 0.1 percent difference with the WKB values in the worst cases when $\ell =1$ and $\mathcal{E}$ is moderate. The usage of the WKB formula is quite efficient here, because it has a very good accuracy for $\ell >n$, while the potentially "unsafe"  WKB mode $\ell=n=0$ is not effected by the tidal force in the used approximation and coincides with the Schwarzschild one.

The low laying quasinormal modes for various values of $\ell$ and $m$ are shown in figs. (\ref{fig2}) and (\ref{fig3}).
The modes with $\ell = m$ are noticeably different from those with  $\ell > m$: while the first are monotonically increasing when $\mathcal{E}$ increases, the second one decreases. Such behavior of the spectrum can be explained as follows:
the height of the potential barrier is lowered by the tidal force for $\ell > m$, which means that it becomes easier penetrable for waves,
leading to smaller real oscillation frequencies and longer living modes. At the same time $\ell = m$ modes are governed by the potentials raised by the tidal force, so that both real and imaginary parts of $\omega$ grows, as $\mathcal{E}$   increases.
The lowered potential for all modes, except the one with the highest azimuthal number $m = \ell$, can be explained by a smaller resultant gravitational attraction near the top of the potential due to the tidal gravity, which is opposite to the gravitational attraction of the black hole.
The non-monotonic "splash" of the imaginary part of $\omega$ in fig. (\ref{fig2}) could probably be explained by approaching the limit of validity $\mathcal{E} \ll 1$.

In the limit of large multipole numbers $\ell \gg m \gg 1$, which is the eikonal regime, the peak of the effective potential is situated at
\begin{equation}
r_{max} = 3 M -\frac{2 M}{3 (2 - 7 M^2 \mathcal{E}) \ell^2} + O(\ell^{-4}).
\end{equation}

Then, the first order WKB formula,
\begin{equation}
\omega = \sqrt{V - i \sqrt{- 2 V_{0}''}\left(n+\frac{1}{2}\right)},
\end{equation}
expanded in powers of $1/\ell$, gives:
\begin{equation}\label{62}
\omega = \frac{\sqrt{1-\frac{7}{2}\mathcal{E} M^2 }}{3 \sqrt{3} M} \left(\ell + \frac{1}{2}\right) - i \left(n+\frac{1}{2}\right) + O(1/\ell).
\end{equation}
In the limit $\mathcal{E} =0$, the above expression reduces to the well-known form for the Schwarzschild black hole \cite{Ferrari}. The quasinormal modes do not depend on $m$ in the first two dominant orders: the dependence on $m$ is in terms
of order $\sim 1/\ell$. The imaginary part of $\omega$ does not depend on $\mathcal{E}$ in the eikonal regime,
which is in concordance with the very slight dependence of the imaginary part on $\mathcal{E}$ in figures (2) and (3).
It is well known that the eikonal formulas similar to (\ref{62}) work very well already for moderate values of $\ell$ and,
in the eikonal regime, a universal behavior for all bosonic fields takes place. Therefore, expression (\ref{62}) is likely to be valid not only for scalar but also for other boson fields in the eikonal regime.

An interesting question is the gravitational stability of the considered system of a torus and a black hole.
The full assessment of the stability issue should be performed by considering the gravitational perturbations of the system, yet, as we can see from the above data, the scalar field shows no unstable modes. A tendency to the instability shows $\ell=m$ modes through the decreasing of the damping rate, which however cannot be extended to sufficiently large values of $\mathcal{E}$ as our effective potential was obtained for the regime of relatively small tidal force.

\section{Conclusions}

The black hole deformed by the surrounding matter and the magnetic field is a background with a relatively
low symmetry, so that one could not expect the separability of variables in either Hamilton-Jacobi or Klein-Gordon equations.
Using a few justified approximations, we have decoupled variables in both equations and
made the problems of particle motion and quasinormal modes solvable with relatively simple analytical and numerical tools.

Initially we have studied motion of charged, massive particles in the equatorial plane.
Two approximations were used here: 1) we considered only a limited spacial region which starts at the event horizon and
finishes at some large distance from the black hole, being still quite far from the place where the tidal force becomes significant; 2) we were limited by relatively small values of the magnetic field and tidal force.
These approximations were stipulated by applicability of the perturbatively obtained Preston-Poisson metric \cite{Preston-Poisson} and
was, therefore, unavoidable.
Consideration of motion only in the equatorial plane was due to the inseparability of the Hamilton-Jacobi equation in the general case.
This limitation is also justified, because, as it was shown in \cite{Galtsov-book} for the Ernst solution, motion in the equatorial plane would be stable against small perturbations perpendicular to the plane, so that the particle which started to move in the equatorial plane, would continue moving in it. However, an analysis of the stability against non-equatorial perturbations, though similar to \cite{Galtsov-book}, is lacking.

Motion of particles is qualitatively different for left and right handed rotations due to the opposite direction of the Lorentz force:
when the signs of the particle's charge and angular momenta coincide (Larmor motion) we have a kind of cyclotron rotation in the magnetic field
"perturbed" by a black hole, while in opposite rotation (anti-Larmor motion), the presence of the black hole is essential.
We have found the energy and angular momentum, the binding energy and the region of radial stability on circular orbits
for the Preston-Poisson space-time. From this we conclude that \emph{the tidal force, as well as the magnetic field, considerably enhances the release of the binding energy and makes the region of stability of circular orbits closer to the black hole.}

The second part is devoted to perturbations of the massless scalar field in the Preston-Poisson background, namely, to the estimations of the characteristic quasinormal modes.
As the decoupling of variables is impossible for the Klein-Gordon equation in the general Preston-Poisson space-time,
we used an additional approximation based on the fact that the low-laying quasinormal modes are "localized" near the peak of the effective potential. Therefore, we performed a kind of averaging of the tidal force by its value at the peak of potential.
\emph{We found a significant decrease in the real oscillation frequency with the tidal force. A simple analytical expression for the frequency has been obtained in the eikonal regime.} Taking into consideration
possible similarities between the quasinormal spectrum of a scalar field with the spectra of gravitational and other long-range neutral boson fields, we neglected the influence of the magnetic field
when studying the quasinormal modes.
Although we managed to estimate the modes with $\ell \geq 1$, the monopole mode is, unfortunately, undistinguishable from its Schwarzschild value in the approximation. All the limitations of our analysis are a price for the simplicity afforded by the approximations employed.
A more accurate numerical investigation would require very time consuming computations and mathematical modeling of the above system.

\acknowledgments{
R. K. acknowledges hospitality of the Centro de Estudios Cientifcos (Valdivia, Chile) where this work was initiated.
At its initial stage the work was partially funded by
the Conicyt grant ACT-91: Southern Theoretical Physics Laboratory (STPLab).
The Centro de Estudios Cientifcos is funded by the Chilean Government
through the Centers of Excellence Base Financing Program of Conicyt.
At the final stage R. K. was supported by the European Commission grant through the Marie Curie International Incoming Contract.
The authors acknowledge A. Zhidenko for useful discussions and and C. J. Pynn for proofreading of the manuscript.
Y-C. L. would like to acknowledge Kostas Kokkotas for financial
support via the German Research Foundation (Deutsche
Forschungsgemeinschaft, DFG) SFB/TR7 on Gravitational Wave Astronomy
and Reinhard Meinel for financial support via the DFG Graduate Academy,
Research training group (1523/1) Quantum and Gravitational field, G3:
Axisymmetric gravitational fields.





\begin{thebibliography}{99}
\bibitem{1} J. L. Han, arXiv:astro-ph/0603512; W. M. Zhang, Y. Lu and S. N. Zhang,
arXiv:astro-ph/0501365; M. Y. Piotrovich, N. A. Silant'ev, Yu. N. Gnedin and
T. M. Natsvlishvili, arXiv:1002.4948 [astro-ph.CO].
\bibitem{2}  R. D. Blandford and R. L. Znajek, Mon. Not. Roy. Astron. Soc. 179 (1977) 433.
\bibitem{Galtsov-book}
  D.~V.~Gal'tsov, ``Particles and fields in the vicinities of the black holes,''
{\it  Moscow University Press, Oct 1986}
\bibitem{Kokkotas:2010zd}
  K.~D.~Kokkotas, R.~A.~Konoplya and A.~Zhidenko,
  Phys.\ Rev.\  D {\bf 83} (2011) 024031
  [arXiv:1011.1843 [gr-qc]].
\bibitem{Preston-Poisson} B.Preston, E. Poisson,   Phys.\ Rev.\  D {\bf 74}, 064010 (2006)
\bibitem{Preston:2006zd}
  B.~Preston and E.~Poisson,
  Phys.\ Rev.\ D {\bf 74}, 064009 (2006).
\bibitem{QNM-collect}
  R.~A.~Konoplya and A.~Zhidenko,
  Rev.\ Mod.\ Phys.\  {\bf 83}, 793 (2011)
  [arXiv:1102.4014 [gr-qc]];
  E.~Berti, V.~Cardoso and A.~O.~Starinets,
  Class.\ Quant.\ Grav.\  {\bf 26}, 163001 (2009);
  K.~D.~Kokkotas and B.~G.~Schmidt,
  Living Rev.\ Rel.\  {\bf 2}, 2 (1999);
  H.~-P.~Nollert,
  Class.\ Quant.\ Grav.\  {\bf 16}, R159 (1999).
\bibitem{cosmology}
  W.~A.~Hiscock,
  Phys.\ Rev.\  {\bf D17}, 1497-1500 (1978);
  D.~A.~Leahy,
  Int.\ J.\ Theor.\ Phys.\  {\bf 21}, 703-753 (1982);
  S.~N.~Guha Thakurta,
  Phys.\ Rev.\  {\bf D21}, 864-966 (1980);
  R.~A.~Konoplya,
  Phys.\ Lett.\ B {\bf 706}, 451 (2012)
  [arXiv:1109.6215 [hep-th]];
  R.~A.~Konoplya and A.~Zhidenko,
  arXiv:1110.2015 [hep-th].
\bibitem{Preti:2009zz}
  G.~Preti,
  Int.\ J.\ Mod.\ Phys.\ D {\bf 18}, 529 (2009).
\bibitem{Preti:2005bu}
  G.~Preti and F.~de Felice,
  Phys.\ Rev.\ D {\bf 71}, 024009 (2005).
\bibitem{Aliev:2002nw}
  A.~N.~Aliev and N.~Ozdemir,
  Mon.\ Not.\ Roy.\ Astron.\ Soc.\  {\bf 336} (2002) 241.
\bibitem{Aliev:1989wx}
  A.~N.~Aliev and D.~V.~Galtsov,
  Sov.\ Phys.\ Usp.\  {\bf 32}, 75 (1989).
\bibitem{Esteban} E. P. Esteban, E. Raamos, Phys. Rev. D 38, 2963 (1988)
\bibitem{Konoplya:2006gg}
  R.~A.~Konoplya,
  Phys.\ Lett.\  B {\bf 644} (2007) 219
  [arXiv:gr-qc/0608066].
\bibitem{Konoplya:2007yy}
  R.~A.~Konoplya, R.~D.~B.~Fontana,
  Phys.\ Lett.\  {\bf B659}, 375-379 (2008).
  [arXiv:0707.1156 [hep-th]].
\bibitem{Konoplya:2008hj}
  R.~A.~Konoplya,
  Phys.\ Lett.\  {\bf B666}, 283-287 (2008).
  [arXiv:0801.0846 [hep-th]].
\bibitem{Konoplya:2006qr}
  R.~A.~Konoplya,
  Phys.\ Rev.\  {\bf D74}, 124015 (2006).
  [gr-qc/0610082].
\bibitem{Kovacs:2011ec}
  Z.~Kovacs, L.~A.~Gergely and M.~Vasuth,
  Phys.\ Rev.\  D {\bf 84} (2011) 024018.
\bibitem{Abdujabbarov:2009az}
  A.~Abdujabbarov and B.~Ahmedov,
  Phys.\ Rev.\  D {\bf 81} (2010) 044022.
\bibitem{Frolov:2010mi}
  V.~P.~Frolov and A.~A.~Shoom,
  Phys.\ Rev.\ D {\bf 82}, 084034 (2010).
\bibitem{Rahimov:2011fv}
  O.~G.~Rahimov, A.~A.~Abdujabbarov and B.~J.~Ahmedov,
  Astrophys.\ Space Sci.\  {\bf 335}, 499 (2011).
\bibitem{Konoplya:2010ak}
  R.~A.~Konoplya,
  Eur.\ Phys.\ J.\  C {\bf 69} (2010) 555
  [arXiv:1002.2818 [hep-th]].
\bibitem{Konoplya:2006rv}
  R.~A.~Konoplya, A.~Zhidenko,
  Phys.\ Lett.\  {\bf B644}, 186-191 (2007).
  [gr-qc/0605082];
  R.~A.~Konoplya, A.~Zhidenko,
  Phys.\ Lett.\  {\bf B648}, 236-239 (2007).
  [hep-th/0611226].
\bibitem{WKB}B.~F.~Schutz and C.~M.~Will Astrophys.\ J.\ Lett {\bf 291} L33 (1985);
S.~Iyer and C.~M.~Will Phys.\ Rev.\  D {\bf 35} 3621 (1987).
\bibitem{WKBorder}
R. A. Konoplya, Phys.\ Rev\ D {\bf 68}, 024018 (2003);
  R.~A.~Konoplya,
  J.\ Phys.\ Stud.\  {\bf 8}, 93 (2004).
\bibitem{WKBuse}
  O.~P.~F.~Piedra and J.~de Oliveira,
  Class.\ Quant.\ Grav.\  {\bf 28}, 085003 (2011).
  O.~P.~F.~Piedra and J.~de Oliveira,
  Class.\ Quant.\ Grav.\  {\bf 28}, 085023 (2011);
  H.~Kodama, R.~A.~Konoplya and A.~Zhidenko,
  Phys.\ Rev.\ D {\bf 81}, 044007 (2010)
  [arXiv:0904.2154 [gr-qc]];
  K.~Lin, J.~Li and N.~Yang,
  Gen.\ Rel.\ Grav.\  {\bf 43}, 1889 (2011);
  K.~Lin,
  Int.\ J.\ Theor.\ Phys.\  {\bf 49}, 2786 (2010);
  K.~A.~Bronnikov, R.~A.~Konoplya and A.~Zhidenko,
  arXiv:1205.2224 [gr-qc].
\bibitem{timedomain}
  C.~Gundlach, R.~H.~Price and J.~Pullin,
  Phys.\ Rev.\  D {\bf 49} (1994) 883;
\bibitem{time-domain-apply}
  E.~Abdalla, R.~A.~Konoplya, C.~Molina,
  Phys.\ Rev.\  {\bf D72}, 084006 (2005)
  [hep-th/0507100];
  R.~A.~Konoplya and A.~Zhidenko,
  Phys.\ Rev.\ Lett.\  {\bf 103}, 161101 (2009)
  [arXiv:0809.2822 [hep-th]].
\bibitem{Ferrari} V. Ferrari, B. Mashhoon, Phys. Rev. Lett. 52, 1361 (1984).
\end{thebibliography}
\end{document}